\documentclass[11pt,letterpaper]{article}

\usepackage{latexsym,multirow}
\usepackage{amssymb,amsmath, bm}
\usepackage{graphicx,subcaption}

\usepackage{enumerate}

\usepackage{natbib}
\usepackage[pdftex, bookmarksopen=true, bookmarksnumbered=true,
pdfstartview=FitH, breaklinks=true, urlbordercolor={0 1 0}, citebordercolor={0 0 1}]{hyperref}
\usepackage{colortbl}

\usepackage{dcolumn}
\newcolumntype{.}{D{.}{.}{-1}}
\newcolumntype{d}[1]{D{.}{.}{#1}}
\usepackage{theorem}
\theoremstyle{plain}
\theoremheaderfont{\scshape}
\newtheorem{assumption}{Assumption}
\newtheorem{example}{Example}

\newtheorem{corollary}{Corollary}

\newtheorem{theorem}{Theorem}

\newtheorem{lemma}{Lemma}

\newcommand{\ind}{\mbox{$\perp\!\!\!\perp$}}

\usepackage{rotating}

\usepackage[compact]{titlesec}

\allowdisplaybreaks

\newcommand\spacingset[1]{\renewcommand{\baselinestretch}%
{#1}\small\normalsize}

\newcommand{\blind}{0}

\newcommand*{\QEDB}{\hfill\ensuremath{\square}}

\def\pr{\textnormal{pr}}    
\def\E{E}
\def\CACE{\tau_c}    
\def\tCACE{\tilde{\tau}_c}    
\def\SP{\textsc{SP}}
\def\SN{\textsc{SN}}

\begin{document} 

\newcommand{\tit}{Measurement errors in the binary instrumental variable model
}
%
%
\spacingset{1.25}

\if0\blind

{\title{\bf\tit}

  \author{ 
  Zhichao Jiang\thanks{Department of Statistics and Department of Government, Harvard University, Cambridge Massachusetts 02138,  Email:
      \href{mailto:zhichao\_jiang@fas.harvard.edu}{zhichao\_jiang@fas.harvard.edu} }   \hspace{.5in} 
      Peng Ding\thanks{Department of Statistics, University of California, Berkeley, California 94720,  \href{mailto:pengdingpku@berkeley.edu}{pengdingpku@berkeley.edu}  }
       }

\date{
\today
}

\maketitle

}\fi

\if1\blind
\title{\bf \tit}

\maketitle
\fi

\pdfbookmark[1]{Title Page}{Title Page}

\thispagestyle{empty}
\setcounter{page}{0}

\begin{abstract}
Instrumental variable methods can identify causal effects even when the treatment and outcome are confounded. We study the problem of imperfect measurements of the binary instrumental variable, treatment or outcome. We first consider non-differential measurement errors, that is, the mis-measured variable does not depend on other variables given its true value. We show that the measurement error of the instrumental variable does not bias the estimate, the measurement error of the treatment biases the estimate away from zero, and the measurement error of the outcome biases the estimate toward zero. Moreover, we derive sharp bounds on the causal effects without additional assumptions. These bounds are informative because they exclude zero. We then consider differential measurement errors, and focus on sensitivity analyses in those settings. \\

\noindent {\bf Keywords:} Complier average causal effect; Misclassification; Noncompliance;  Sensitivity analysis; Sharp bound
\end{abstract}


\clearpage
\spacingset{1.5}

\section{Introduction}

Instrumental variable methods are powerful tools for causal inference with unmeasured treatment-outcome confounding.
\citet{angrist1996identification} use potential outcomes to clarify the role of a binary instrumental variable in identifying causal effects. They show that the classic two-stage least squares estimator is consistent for the complier average causal effect under the monotonicity and exclusion restriction assumptions. 


Measurement error is common in empirical research, which is also called misclassification for discrete variables.
\citet{black2003measurement} study the return of a possibly misreported education status. \citet{boatman2017estimating} study the effect of a self-reported smoking status. In those settings, the treatments are endogenous and mismeasured. \citet{chalak2017instrumental} considers the measurement error of an instrumental variable. \citet{pierce2012effect} consider a continuous treatment and either a continuous or a binary outcome with measurement errors. The existing literature often relies on modeling assumptions \citep{schennach2007instrumental, pierce2012effect}, auxiliary information \citep{black2003measurement, kuroki2014measurement,  chalak2017instrumental,boatman2017estimating}, or repeated measurements of the unobserved variables \citep{battistin2014misreported}.

With binary variables, we study all possible scenarios of measurement errors of the instrumental variable, treatment and outcome. Under non-differential measurement errors, we show that the measurement error of the instrumental variable does not result in bias, the measurement error of the treatment moves the estimate away from zero, the measurement error of the outcome moves the estimate toward zero. 
This differs from the result for the total effect \citep{bross1954misclassification} where measurement errors of the treatment and outcome both move the estimate toward zero. 

For non-differential measurement errors, we focus on qualitative analysis and nonparametric bounds. For differential measurement errors, we focus on sensitivity analysis. In both cases, we do not impose modeling assumptions or require auxiliary information.

\section{Notation and assumptions for the instrumental variable estimation}
For unit $i$, let $Z_i$ denote the treatment assigned, $D_i$ the treatment received, and $Y_i$ the outcome. Assume that $ (Z_i,D_i, Y_i)$ are all binary taking values in $\{0,1\}$. We ignore pretreatment covariates without loss of generality, because all the results hold within strata of covariates. We use potential outcomes to define causal effects. Define the potential values of the treatment received and the outcome as $D_{zi}$ and  $Y_{zi}$ if unit $i$ were assigned to treatment arm $z$ ($z=0, 1$). The observed values are $D_i = Z_iD_{1i} + (1-Z_i)D_{0i}$ and $Y_i  = Z_iY_{1i} + (1 - Z_i)Y_{0i}$. \citet{angrist1996identification} classify the units into four latent strata based on the joint values of $(D_{1i}, D_{0i} ) $. They define $U_i=a$ if $(D_{1i},  D_{0i} ) =(1,1)$, $U_i=n$ if $(D_{1i},  D_{0i} ) =(0,0)$, $U_i=c$ if $(D_{1i},  D_{0i} ) =(1,0)$, and $U_i=d$ if $(D_{1i},  D_{0i} ) =(0,1)$. The stratum with $U_i=c$ consists of compliers.
For notational simplicity, we drop the subscript $i$. We invoke the following assumption for the instrumental variable model. 

\begin{assumption}
\label{asm:iv}
Under the instrumental variable model, 
(a) $Z \ind (Y_1, Y_0, D_1, D_0)$, (b) $D_1 \geq D_0$, and (c) $\pr(Y_1=1 \mid U=u) =\pr(Y_0=1 \mid U=u)$ for $u=a$ and $n$.
\end{assumption}

Assumption~\ref{asm:iv}(a) holds  in randomized experiments.  Assumption~\ref{asm:iv}(b) means that the treatment assigned has a monotonic effect on the treatment received for all units, which rules out the latent strata $U=d.$ Assumption~\ref{asm:iv}(c) implies that the treatment assigned affects the outcome only through the treatment received,  which is called exclusion restriction.

Define $\textsc{RD}_{R\mid Q}= \pr(R=1 \mid Q=1)-\pr(R=1 \mid Q=0)$ as the risk difference of $Q$ on $R$. For example, $\textsc{RD}_{YD\mid (1-Z)}= \pr(Y=1,D=1 \mid Z=0)-\pr(Y=1,D=1 \mid Z=1)$. \citet{angrist1996identification} show that the complier average causal effect 
\begin{eqnarray*}
\CACE
\equiv \E(Y_1-Y_0\mid U=c)
=\frac{\pr(Y=1 \mid Z=1)-\pr(Y=1 \mid Z=0)}{\pr(D=1 \mid Z=1)-\pr(D=1 \mid Z=0)}
=\frac{\textsc{RD}_{Y\mid Z}}{\textsc{RD}_{D\mid Z}}
\end{eqnarray*}
can be identified by the ratio of the risk differences of $Z$ on $Y$ and $D$ if $\textsc{RD}_{D\mid Z}\neq 0.$

\section{Non-differential measurement errors}\label{sec::nondiffmeasure}
Let $(Z',D',Y')$ denote the possibly mismeasured values of $(Z,D,Y)$. Without the true variables, we use the naive estimator based on the observed variables to estimate $\CACE$:
\begin{eqnarray*}
\CACE'
\equiv \frac{\pr(Y'=1 \mid Z'=1)-\pr(Y'=1 \mid Z'=0)}{\pr(D'=1 \mid Z'=1)-\pr(D'=1 \mid Z'=0)}= \frac{\textsc{RD}_{Y'\mid Z'}}{\textsc{RD}_{D'\mid Z'}}.
\end{eqnarray*}

\begin{assumption}
\label{asm:nondif}
All measurement errors are  non-differential:
$\pr(D' \mid D, Z',Z,Y,Y')=\pr(D' \mid D)$, $\pr(Y' \mid Y, Z,Z',D,D')=\pr(Y' \mid Y)$, and $\pr(Z' \mid Y, Y',Z,D,D')=\pr(Z' \mid Z) .$
\end{assumption}

Under Assumption \ref{asm:nondif}, the measurements of the variables do not depend on other variables conditional on the unobserved true variables. We use the sensitivities and specificities to characterize the non-differential measurement errors: 
\begin{align*}
 \textsc{SN}_D&=\pr(D'=1 \mid D=1),\quad & \textsc{SP}_D&=\pr(D'=0 \mid D=0),\quad  & r_D &= \textsc{SN}_D+\textsc{SP}_D-1 \leq 1, \\ 
 \textsc{SN}_Y&=\pr(Y'=1 \mid Y=1)  , \quad &\textsc{SP}_Y&=\pr(Y'=0 \mid Y=0),\quad  &r_Y &= \textsc{SN}_Y+\textsc{SP}_Y-1 \leq 1.
\end{align*}
Without measurement errors, $r_D = r_Y = 1.$
Assume $r_D >0$ and $r_Y >0$, which means that the observed variable is informative for the true variable, i.e., the observed variable is more likely to be 1 if the true variable is 1 rather than 0. We state a simple relationship between $\CACE$ and $\CACE'$.
\begin{theorem}
\label{thm:cace}
Under Assumptions~\ref{asm:iv} and~\ref{asm:nondif}, 
$\CACE = \CACE' \times r_D/r_Y$.
\end{theorem}

Theorem \ref{thm:cace} shows that measurement errors of $Z$, $D$ and $Y$ have different consequences. The measurement error of $Z$ does not bias the estimate. The measurement error of $D$ biases the estimate away from zero. The measurement error of $Y$ biases the estimate toward zero. In contrast, measurement errors of the treatment and outcome both bias the estimate toward zero in the total effect estimation \citep{bross1954misclassification}.  

Moreover, the measurement errors of $D$ and $Y$ have mutually independent influences on the estimation of $\CACE$.
Theorem \ref{thm:cace} also shows that $\CACE$ and $\CACE'$ have the same sign when $r_D>0$ and $r_Y > 0$.

\section{Bounds on $\CACE$ with non-differential measurement errors}
\label{sec::nondiff}

When $D$ or $Y$ is non-differentially mismeasured, we can identify $\CACE$ if we know $r_D$ and $r_Y$. Without knowing them, we cannot identify $\CACE$. Fortunately, the observed data still provide some information about $\CACE$. We can derive its sharp bounds based on the joint distribution of the observed data.    We first introduce a lemma.

\begin{lemma}
\label{lem:bound}
Define $\textsc{SN}'_Z= \pr(Z=1\mid Z'=1)$ and $\textsc{SP}'_Z= \pr(Z=0\mid Z'=0)$.
Under Assumption 1, given the values of $(\textsc{SN}'_Z,\textsc{SP}'_Z,\textsc{SN}_D,\textsc{SP}_D,\textsc{SN}_Y,\textsc{SP}_Y)$, there is a one-to-one mapping between the set $\{\pr(Z=z), \pr(U=u),\pr(Y_z=1\mid U=u) : z=0,1; u=a,n,c\}$ and the set $\{  \pr(Z'=z', D' = d', Y'=y'): z', d', y' = 0,1  \}$. 
\end{lemma}

Lemma~\ref{lem:bound} allows for simultaneous measurement errors of more than one elements of $(Y, Z, D)$.
From Lemma~\ref{lem:bound}, given the sensitivities and specificities, we can recover the joint distribution of $(Y_z,U,Z)$ for $z=0,1$. Conversely, the conditions $\{  0\leq \pr(Z=z) \leq 1,  0\leq \pr(U=u) \leq 1, 0\leq \pr(Y_z=1\mid U=u) \leq 1 : z=0,1;u=a,n,c\}$ induce sharp bounds on the sensitivities and specificities, which further induce sharp bounds on $\CACE$. This is a general strategy that we use to derive sharp bounds on $\CACE$.

First, we discuss the measurement error of $Y$. 
\begin{theorem}
\label{thm:bound:Y}
Suppose that $\CACE '\geq 0$ and only  $Y$ is mismeasured with $r_Y>0$.
Under Assumptions~\ref{asm:iv} and~\ref{asm:nondif}, the sharp bounds are  
$\textsc{SN}_Y \geq M_Y$, $\textsc{SP}_Y \geq 1- N_Y$, and $\CACE ' \leq \CACE  \leq \CACE '/(M_Y- N_Y)$,
where $M_Y$ and $N_Y$ are the maximum and minimum values of the set
\begin{eqnarray*}
\left \{\pr(Y'=1\mid D=0, Z=1),\pr(Y'=1 \mid D=1, Z=0), \frac{\textsc{RD}_{Y'D\mid Z}}{\textsc{RD}_{D\mid Z}}, \frac{\textsc{RD}_{Y'(1-D)\mid (1-Z)}}{\textsc{RD}_{D\mid Z}}\right\}.
\end{eqnarray*}
\end{theorem}

We can obtain the bounds under $\CACE ' <0$ by replacing $Y$ with $1-Y$ and $Y'$ with $1-Y'$ in Theorem \ref{thm:bound:Y}. Thus, we only consider $\CACE ' \geq 0$ in Theorem \ref{thm:bound:Y} and the theorems in later parts of the paper.
In Theorem \ref{thm:bound:Y}, the lower bounds on $\textsc{SN}_Y $ and $\textsc{SP}_Y$ must be smaller than or equal to $1$, i.e., $M_Y \leq 1$ and $1- N_Y \leq 1$. These two inequalities further imply the following corollary on the testable conditions of the instrumental variable model with the measurement error of $Y.$

\begin{corollary}
\label{cor:testY}
Suppose that only $Y$ is mismeasured with $r_Y>0$. 
Under Assumptions~\ref{asm:iv} and~\ref{asm:nondif},  
\begin{eqnarray*}
\pr(Y'=y,D=1 \mid Z=1) &\geq& \pr(Y'=y, D=1 \mid Z=0), \quad (y=0,1),\\
\pr(Y'=y,D=0 \mid Z=0) &\geq& \pr(Y'=y, D=0 \mid Z=1), \quad (y=0,1).
\end{eqnarray*}
\end{corollary}

The conditions in Corollary \ref{cor:testY} are all testable with observed data $(Z,D,Y')$, and they are the same under $\CACE ' \geq 0$ and $\CACE '<0$. \citet{balke1997bounds} derive the same conditions as in Corollary \ref{cor:testY} without the measurement error of $Y$. \citet{wang2017falsification} propose statistical tests for these conditions. From Corollary \ref{cor:testY}, the non-differential measurement error of $Y$ does not weaken the testable conditions of the binary instrumental variable model.

Second, we discuss the measurement error of $D$.

\begin{theorem}
\label{thm:bound:D}
Suppose that $\CACE ' \geq 0$ and only $D$ is mismeasured with $r_D>0$. Under
Assumptions~\ref{asm:iv} and~\ref{asm:nondif}, the sharp bound are 
$
 M_D\leq \textsc{SN}_D \leq U_D ,$
 $
 1-N_D \leq \textsc{SP}_D \leq 1-V_D,
$
and
$ 
 \CACE ' \times (M_D - N_D) \leq \CACE  \leq  \CACE ' \times (U_D - V_D),
$
where 
\begin{eqnarray*}
M_D&=& 
\max \left\{ \max_{z=0,1}  \pr(D'=1\mid Z=z),  \max_{y=0,1}  \pr(D'=1 \mid Y=y,Z=1),\frac{\textsc{RD}_{(1-Y)D'\mid (1-Z)}}{\textsc{RD}_{Y\mid Z}}\right\} ,\\ 
N_D&=&
\min \left\{\min_{z=0,1}  \pr (D'=1\mid Z=z),\min_{y=0,1}  \pr (D'=1 \mid Y=y,Z=0),\frac{\textsc{RD}_{YD'\mid Z}}{\textsc{RD}_{Y\mid Z}}\right\} ,\\
U_D &=& \min\left \{1,\frac{\textsc{RD}_{YD'\mid Z}}{\textsc{RD}_{Y\mid Z}}\right\},\quad
V_D = \max\left \{0,\frac{\textsc{RD}_{(1-Y)D'\mid (1-Z)}}{\textsc{RD}_{Y\mid Z}}\right\} . 
\end{eqnarray*}
\end{theorem}

With a mis-measured $D$, \citet{ura2018heterogeneous} derives sharp bounds with and without Assumption \ref{asm:nondif}, respectively. The former bounds are equivalent to ours, but the latter bounds are wider. In Theorem \ref{thm:bound:D}, the lower bounds on $\textsc{SN}_D$ and $\textsc{SP}_D$ must be smaller than or equal to their upper bounds. This further implies the following corollary on the testable conditions of the binary instrumental variable model with the measurement error of $D.$

\begin{corollary}
\label{cor:testD}
Suppose that $\CACE ' \geq 0$ and only $D$ is mismeasured with $r_D>0$. 
Under Assumptions~\ref{asm:iv} and~\ref{asm:nondif}, 
\begin{eqnarray}
\pr(Y=1,D'=1 \mid Z=1) &\geq&  \pr(Y=1, D'=1 \mid Z=0), \label{eq::testableD1}  \\
\pr(Y=0,D'=0 \mid Z=0) &\geq &\pr(Y=0, D'=0 \mid Z=1), \nonumber  \\
\pr(D'=1 \mid Y=y,Z=1) &\leq  & \textsc{RD}_{YD'\mid Z} / \textsc{RD}_{Y\mid Z},  \quad  \hspace{1.23cm}
  (y=0,1), \nonumber \\
 \pr(D'=1 \mid Y=y,Z=0) &\geq  &  \textsc{RD}_{(1-Y)D'\mid (1-Z)} / \textsc{RD}_{Y\mid Z} , \quad  (y=0,1). \nonumber 
\end{eqnarray}
\end{corollary}

We can obtain the conditions under $\CACE ' < 0$ by replacing $Y$ with $1-Y$. 
In the Supplementary material, we show that the conditions in Corollary~\ref{cor:testD} are weaker than those in \citet{balke1997bounds}.
Thus, the non-differential measurement error of $D$ weakens the testable conditions of the binary instrumental variable model.

It is complicated to obtain closed-form bounds under simultaneous measurement errors of more than one elements of $(Z, D, Y)$. In those cases, we can numerically calculate the sharp bounds on $\CACE$ with details in the Supplementary material.

\section{Results under strong monotonicity}
\label{sec::mono}

Sometimes, units in the control group have no access to the treatment. It is called the one-sided noncompliance problem with the following assumption.

\begin{assumption}
\label{asm:str}
 For all individual $i$, 
$D_{0i}=0$.
\end{assumption}

Under strong monotonicity, we have only two strata with $U=c$ and $U=n$. Theorem \ref{thm:cace} still holds. 
Moreover, strong monotonicity sharpens the bounds in \S \ref{sec::nondiff}.

First, we consider the measurement error of $Y.$ We have 
\begin{eqnarray*}
\CACE ' =\left\{  \pr(Y'=1 \mid Z=1)-\pr(Y'=1 \mid Z=0) \right\} / \pr(D=1 \mid Z=1) , \quad \CACE =\CACE '/r_Y.
\end{eqnarray*}

\begin{theorem}
\label{thm:bound:str:Y}
Suppose that  $\CACE ' \geq 0$ and only $Y$ is mismeasured with $r_Y>0$.
Under Assumptions~\ref{asm:iv}--\ref{asm:str}, the sharp bounds are
$\SP_Y \geq  1- N_Y^{\textup{m}}$, $\SN_Y \geq M_Y^{\textup{m}}$, and $\CACE ' \leq \CACE  \leq \CACE '/ (M_Y^{\textup{m}} - N_Y^{\textup{m}})$, where
\begin{eqnarray*}
N_Y^{\textup{m}} &=& \min \{\pr(Y'=1\mid D=0,Z=1), \pr(Y'=1 \mid D=1,Z=1)-\CACE '\},\\
M_Y^{\textup{m}} &=&  \max \{\pr(Y'=1\mid D=0,Z=1), \pr(Y'=1\mid D=1,Z=1)\}. 
\end{eqnarray*}
\end{theorem}

Second, we consider the measurement error of $D.$ Subtle issues arise. When $D$ is mismeasured, $\pr(D'=0 \mid D=0, Z=0)=1$ is known, and $\pr(D'=1 \mid D=1, Z=0)$ is not well defined. Thus, Assumption \ref{asm:nondif} of non-differential measurement error is implausible. We need modifications. Define
\begin{eqnarray*}
&\textsc{SN}_D^1&= \pr(D'=1 \mid D=1, Z=1), \quad
\textsc{SP}_D^1= \pr(D'=0 \mid D=0, Z=1) 
\end{eqnarray*}
as the sensitivity and specificity conditional on $Z=1$. We have
\begin{eqnarray*}
\CACE ' = \textsc{RD}_{Y\mid Z} /  \left\{   \pr(D'=1 \mid Z=1)-(1-\textsc{SP}_D^1) \right\},\quad 
\CACE  =  \CACE ' \times (\textsc{SN}_D^1+\textsc{SP}_D^1-1) .
\end{eqnarray*}

\begin{theorem}
\label{thm:bound:str:D} 
Suppose that $\CACE ' \geq 0$,  only $D$ is mismeasured, and 
\begin{equation}
\pr(D'=1 \mid Y=1,Z=1) \geq \pr(D'=1 \mid Y=0,Z=1).
\label{eq::conditionD}
\end{equation}
Under Assumptions~\ref{asm:iv} and~\ref{asm:str}, the sharp bounds are
\begin{eqnarray*}
&&\textsc{SP}_D^1 \geq 1- \pr(D'=1 \mid Y=0,Z=1),  \quad \textsc{SN}_D^1 \geq \pr(D'=1 \mid Y=1,Z=1),\\
&&   \textsc{SN}_D^1 \leq \left\{  \pr(Y=1,D'=1 \mid Z=1)-(1-\textsc{SP}_D^1)\times  \pr(Y=1\mid Z=0) \right\} / \textsc{RD}_{Y\mid Z}  ,\\
&& \pr(D'=1 \mid Y=1, Z=1)\times \textsc{RD}_{Y\mid Z} / \pr(D'=1 \mid Z=1)  \leq \CACE  \leq 1 .
\end{eqnarray*}
\end{theorem}

Unlike Theorems~\ref{thm:bound:Y}--\ref{thm:bound:str:Y},
the upper bound on $ \textsc{SN}_D^1$ depends on  $\textsc{SP}_D^1$ in Theorem~\ref{thm:bound:str:D}. 
The condition in \eqref{eq::conditionD} is not necessary for obtaining the bounds, but it helps to simplify the expression of the bounds.
It holds in our applications in \S \ref{sec::illustration}. We give the bounds on $\CACE$ without \eqref{eq::conditionD} in the Supplementary material. The upper bound on $\CACE$ is not informative in Theorem~\ref{thm:bound:str:D}, but,
fortunately, we are more interested in the lower bound in this case.

It is complicated to obtain closed-form bounds under simultaneous measurement errors of more than one elements of $(Z, D, Y)$. In those cases, we can numerically calculate the sharp bounds with more details in the Supplementary material.

\section{Sensitivity analysis formulas under differential measurement errors}
\label{sec::sensitivityanalysis}

Non-differential measurement error is not plausible in some cases.
\S \ref{sec::mono} shows that under strong monotonicity, the measurement error of $D$ cannot be non-differential because it depends on $Z$ in general.  In this section, we consider differential measurement errors of $D$ and $Y$ without requiring strong monotonicity. We do not consider the differential measurement error of $Z$, because the measurement of $Z$ often precedes $(D, Y)$ and its measurement error is unlikely to depend on later variables.

We first consider the differential measurement error of $Y$.
 
\begin{theorem}
\label{thm:diff:Y}
 Suppose that only $Y$ is mismeasured.
Define
\begin{align}
\textsc{SN}_Y^1&= \pr(Y'=1 \mid Y=1, Z=1), \quad &\textsc{SN}_Y^0&= \pr(Y'=1 \mid Y=1, Z=0),  \label{eq::misY1} \\
\textsc{SP}_Y^1&= \pr(Y'=0 \mid Y=0, Z=1), \quad &\textsc{SP}_Y^0&= \pr(Y'=0 \mid Y=0, Z=0). \label{eq::misY2}
\end{align}
Under Assumption~\ref{asm:iv},
\begin{eqnarray*}
\CACE = \left\{ \frac{\pr(Y'=1 \mid Z=1)-(1-\textsc{SP}_Y^1)}{\textsc{SN}_Y^1+\textsc{SP}_Y^1-1}-\frac{\pr(Y'=1 \mid Z=0)-(1-\textsc{SP}_Y^0)}{\textsc{SN}_Y^0+\textsc{SP}_Y^0-1}\right\} \bigg/ \textsc{RD}_{D\mid Z}.
\end{eqnarray*}
\end{theorem}

Theorem \ref{thm:diff:Y} allows the measurement error of $Y$ to depend on $D$, but the formula of $\CACE$  only needs
the sensitivities and specificities  in \eqref{eq::misY1} and \eqref{eq::misY2} conditional on $(Z, Y)$.
It is possible that $\CACE'$ is positive but $\CACE$ is negative. For example, if  
$
\textsc{SN}_Y^1+\textsc{SP}_Y^1=\textsc{SN}_Y^0+\textsc{SP}_Y^0>1
$
and
$ \textsc{SP}_Y^0-\textsc{SP}_Y^1>\textsc{RD}_{Y'\mid Z},
$
then $\CACE$ and $\CACE'$ have different signs.

We then consider the differential measurement error of $D$.

\begin{theorem}
\label{thm:diff:D}
 Suppose that only $D$ is mismeasured.
Define 
\begin{align}
\textsc{SN}_D^1&= \pr(D'=1 \mid D=1, Z=1), \quad &\textsc{SN}_D^0&= \pr(D'=1 \mid D=1, Z=0),  \label{eq::misD1}\\
\textsc{SP}_D^1&= \pr(D'=0 \mid D=0, Z=1), \quad &\textsc{SP}_D^0&= \pr(D'=0 \mid D=0, Z=0). \label{eq::misD2}
\end{align}
Under Assumption~\ref{asm:iv},
\begin{eqnarray*}
\CACE =\textsc{RD}_{Y\mid Z} \bigg /\left\{\frac{\pr(D'=1 \mid Z=1)-(1-\textsc{SP}_D^1)}{\textsc{SN}_D^1+\textsc{SP}_D^1-1}-\frac{\pr(D'=1 \mid Z=0)-(1-\textsc{SP}_D^0)}{\textsc{SN}_D^0+\textsc{SP}_D^0-1}\right\}.
\end{eqnarray*}
\end{theorem}

Theorem \ref{thm:diff:D} allows the measurement error of $D$ to depend on $Y$, but the formula of $\CACE$ only needs 
the sensitivities and specificities \eqref{eq::misD1} and \eqref{eq::misD2} conditional on $Z$.
Similar to the discussion after Theorem \ref{thm:diff:Y}, it is possible that $\CACE'$  and $\CACE$  have different signs.

Based on Theorems~\ref{thm:diff:Y} and~\ref{thm:diff:D}, if we know or can consistently estimate the sensitivities and specificities in \eqref{eq::misY1}--\eqref{eq::misD2},  then we can consistently estimate $\CACE$; if we only know the ranges of the sensitivities and specificities, then we can obtain bounds on $\CACE$. 

For simultaneous differential measurement errors of $D$ and $Y$, the formula of $\CACE$ depends on too many sensitivity and specificity parameters. Thus we omit the discussion.  
 
\section{Illustrations}\label{sec::illustration}

We give three examples and present the data in the Supplementary material.

\begin{example}\label{eg::1}
\citet{improve2014endovascular} assess the effectiveness of the emergency endovascular versus the open surgical repair strategies for patients with a clinical diagnosis of ruptured aortic aneurism. Patients are randomized to either the emergency endovascular or the open repair strategy. The primary outcome is the survival status after 30 days. Let $Z$ be the treatment assigned, with $Z=1$ for the endovascular strategy and $Z=0$ for the open repair. Let $D$ be the treatment received. 
Let $Y$ be the survival status, with $Y=1$ for dead, and $Y=0$ for alive. 
If none of the variables are mismeasured, then the estimate of $\CACE $ is $0.131$ with 95\% confidence interval $(-0.036, 0.298)$ including $0$.
If only $Y$ is non-differentially mismeasured, then
$0.382 \leq \textsc{SP}_Y \leq 1$,   $0.759 \leq \textsc{SN}_Y \leq 1$,  $0.141 \leq r_Y \leq 1$, and thus
$0.131 \leq \CACE  \leq 0.928$ from Theorem \ref{thm:bound:Y}.  
If only $D$ is non-differentially mismeasured, then
$0.658 \leq \textsc{SN}_D \leq 1$,   $0.908 \leq \textsc{SP}_D \leq 1$, $0.566 \leq r_D \leq 1$, and thus
$0.074 \leq \CACE  \leq 0.131$ from Theorem \ref{thm:bound:D}. 
\end{example}

\begin{example}\label{eg::2}
In \citet{hirano2000assessing}, physicians are randomly selected to receive a letter encouraging them to inoculate patients at risk for flu. The treatment is the actual flu shot, and the outcome is an indicator for flu-related hospital visits. However, some patients do not comply with their assignments. Let $Z_i$ be the indicator of encouragement to receive the flu shot, with $Z=1$ if the physician receives the encouragement letter, and  $Z=0$ otherwise. Let $D$ be the treatment received.
Let $Y$ be the outcome, with $Y=0$ if for  a flu-related hospitalization during the winter, and $Y=1$ otherwise. 
If none of the variables are mismeasured, then the estimate of $\CACE $ is $0.116$ with 95\% confidence interval $(-0.061, 0.293)$ including $0$.
If only $Y$ is non-differentially mismeasured, then from Theorem \ref{thm:bound:Y},  $\textsc{SP}_Y \geq 1.004 > 1$, and thus the assumptions of the instrumental variable  do not hold.  
If only $D$ is non-differentially mismeasured, then from Theorem \ref{thm:bound:D}, $\textsc{SN}_D \geq 8.676 > 1$, and thus the assumptions of the instrumental variable 
  do not hold either. We reject the testable condition \eqref{eq::testableD1} required by both Corollaries~\ref{cor:testY} and  \ref{cor:testD} with $p$-value smaller than $10^{-9}$.
   As a result, the non-differential measurement error of $D$ or $Y$ cannot explain the violation of the instrumental variable assumptions in this example.
\end{example}

\begin{example}\label{eg::3}
\citet{sommer1991estimating} study the effect of vitamin A supplements
on the infant mortality in Indonesia. The vitamin supplements are randomly assigned to
villages, but some of the individuals in villages assigned to the treatment group do not
receive them. Strong monotonicity holds, because the individuals assigned to the control group have no access to the supplements. Let $Y$ denote a binary outcome, with $Y=1$ if the infant survives to twelve months, and $Y=0$ otherwise. Let $Z$ denote the indicator of assignment to the supplements. Let $D$ denote the actual receipt of the supplements. If none of the variables are mismeasured, then the estimate of $\CACE $ is $0.003$ with 95\% confidence interval $(0.001, 0.005)$ excluding $0$.
If only $Y$ is non-differentially mismeasured, then $\textsc{SP}_Y \geq 0.014$, $\textsc{SN}_Y \geq 0.999$, and thus
$0.003 \leq \CACE  \leq 0.252$ from Theorem \ref{thm:bound:str:Y}. The 95\% confidence interval is $(0.001,1)$.
If only $D$ is non-differentially mismeasured, then $\textsc{SP}^1_D \geq 0.739$, $\textsc{SN}^1_D \geq 0.802$, and thus
$0.003 \leq \CACE  \leq 1$ from Theorem \ref{thm:bound:str:D}.  The 95\% confidence interval is $(-1\times 10^{-5},1)$. 
In the Supplementary material, we give the details for constructing confidence intervals for $\CACE$ based on its sharp bounds.
\end{example} 

In Examples \ref{eg::1} and \ref{eg::3}, the upper bounds on $\CACE$ are too large to be informative, but fortunately, the lower bounds are of more interest in these applications.

\section{Discussion}\label{sec::discussion}

\subsection{Further comments on the measurement errors of $Z$}

If only $Z$ is mismeasured and the measurement error is non-differential, then $\textsc{RD}_{D\mid Z'}= r_Z' \times \textsc{RD}_{D\mid Z}$ where $r_Z' = \textsc{SN}'_Z+\textsc{SP}'_Z-1$ with $\textsc{SN}'_Z$ and $\textsc{SP}'_Z$ defined in Lemma \ref{lem:bound}. If $r_Z' $ and $\textsc{RD}_{D\mid Z}$ are both constants that do not shrink to zero as the sample size $n$ increases, then $\textsc{RD}_{D\mid Z'}$ does not shrink to zero either. In this case, measurement error of $Z$ does not cause the weak instrumental variable problem \citep{nelson1990distribution,staiger1997instrumental}. Theorem~\ref{thm:cace} shows that  the non-differential measurement error of  $Z$ does not affect the large-sample limit of the naive estimator. We further show in the Supplementary material that it does not affect the asymptotic variance of the naive estimator either.

Nevertheless, in finite samples, the measurement error of $Z$ does result in smaller estimate for $\textsc{RD}_{D\mid Z'}$. If we consider the asymptotic regime that $ r_Z' = o(n^{-\alpha})$ for some $\alpha >0$, then it is possible to have the weak instrumental variable problem. In this case, we need tools that are tailored to weak instrumental variables \citep{nelson1990distribution,staiger1997instrumental}.


Practitioners sometimes dichotomize a continuous instrumental variable $Z$ into a binary one based on the median or other quantiles. The dichotomized variable based on other quantiles are measurement errors of the dichotomized variable based on the median. However, these measurement errors are differential and thus our results in \S\ref{sec::nondiffmeasure} and \S\ref{sec::nondiff} are not applicable.


\subsection{Further commments on the measurement errors of $D$}

We discussed binary $D$.  If we dichotomize a discrete $D \in \{0, 1, \ldots,J\}$ at $k$, i.e., $D'=1(D  \geq  k)$, then we can define two-stage least squares estimators based on $D$ and $D'$:
$$
\tau_{\text{2sls}} = 
\frac{  \E(Y\mid Z=1)-\E(Y\mid Z=0) }{ \E(D \mid Z=1)- \E(D \mid Z=0) } ,\quad
\tau_{\text{2sls}} '  = 
\frac{  \E(Y\mid Z=1)-\E(Y\mid Z=0) }{ \E(D' \mid Z=1)- \E(D' \mid Z=0) } .
$$
\citet{angrist1995two} show that $\tau_{\text{2sls}}$
is a weighted average of some subgroup causal effects.
Analogous to Theorem \ref{thm:cace}, we show in the Supplementary material that $ \tau_{\text{2sls}} = \tau_{\text{2sls}} '  \times w_k  $, where 
$w_k = \pr(D_1  \geq k>D_0) / \sum_{j=1}^J \pr(D_1  \geq j>D_0) \in [0,1]$ if Assumptions \ref{asm:iv}(a) and (b) hold. Therefore, the  dichotomization biases  the estimate away from zero.

\subsection{Further comments on the measurement errors of $Y$}


For a continuous outcome,  it is common to assume that the measurement error of $Y$ is additive and non-differential, i.e., $Y'=Y+U$, where $U$ is the error term with mean zero. If the binary $Z$ and $D$ are non-differentially mismeasured as in Assumption \ref{asm:nondif}, then $\CACE = \CACE ' \times r_D$. In this case, the measurement error of $Y$ does not bias the estimate for $\CACE$. 

\newpage
\appendix

\setcounter{equation}{0}
\setcounter{figure}{0}
\setcounter{theorem}{0}
\setcounter{lemma}{0}
\setcounter{section}{0}
\renewcommand {\theequation} {S\arabic{equation}}
\renewcommand {\thefigure} {S\arabic{figure}}
\renewcommand {\thetheorem} {S\arabic{theorem}}
\renewcommand {\thelemma} {S\arabic{lemma}}
\renewcommand {\thesection} {S\arabic{section}}

\begin{center}
  \LARGE {\bf Supplementary Material}
\end{center}

The supplementary material contains six sections \S\S \ref{sec::prooftheorems}--\ref{sec::otherresults} corresponding to \S\S \ref{sec::nondiffmeasure}--\ref{sec::discussion}.

\S\ref{sec::prooftheorems} gives the proof for Theorem \ref{thm:cace} in \S \ref{sec::nondiffmeasure}. 

\S\ref{sec::boundsCACE} gives proofs for Lemma \ref{lem:bound}, Theorems \ref{thm:bound:Y} and \ref{thm:bound:D}, and Corollaries \ref{cor:testY} and \ref{cor:testD} in \S \ref{sec::nondiff}, and details of computing bounds in more complicated cases.

\S\ref{sec::strongmono} gives the proofs of Theorems \ref{thm:bound:str:Y} and \ref{thm:bound:str:D} in \S \ref{sec::mono}, and details of computing bounds in more complicated cases.

\S\ref{sec::differentialmeasurementerror} gives the proofs of Theorems \ref{thm:diff:Y} and \ref{thm:diff:D} in \S \ref{sec::sensitivityanalysis}.

\S\ref{sec::illustrationdetails} gives more details for \S \ref{sec::illustration}, including the data and a method for constructing confidence intervals for $\CACE$ based on its bounds. 

\S\ref{sec::otherresults} gives additional results summarized in \S \ref{sec::discussion}.

\section{Proof of Theorem \ref{thm:cace}}\label{sec::prooftheorems}

\begin{lemma}
\label{lem:error}
Suppose that $S,S',Q,Q'$ are binary variables, and 
$$\pr(S'=s \mid S=s,Q',Q)=\pr(S'=s \mid S=s), \quad \pr(S=s \mid S'=s)=a_s  \quad  (s=0,1),$$ and 
$$\pr(Q'=q \mid Q=q,S',S)= \pr(Q'=q \mid Q=q)=b_q  \quad (q=0,1).$$
Then
\begin{eqnarray*}
\pr(Q=1 \mid S=1) &=&\frac{a_0\pr(Q'=1\mid S'=1)-(1-a_1)\pr(Q'=1\mid S'=0)}{(a_1+a_0-1)(b_1+b_0-1)}
-\frac{1-b_0}{b_1+b_0-1},\\
\pr(Q=1 \mid S=0) &=&\frac{a_1\pr(Q'=1\mid S'=0)-(1-a_0)\pr(Q'=1\mid S'=1)}{(a_1+a_0-1)(b_1+b_0-1)}
-\frac{1-b_0}{b_1+b_0-1},
\end{eqnarray*}
and
\begin{equation}
\label{eq::bross}
\textsc{RD}_{Q'\mid S'} 
= (a_1+a_0-1)(b_1+b_0-1)\textsc{RD}_{Q\mid S}.
\end{equation}
\end{lemma}

The identity \eqref{eq::bross} corroborates \citet{bross1954misclassification}'s result that non-differential measurement error of a binary treatment or outcome biases the estimate of the total effect toward zero if $| a_1+a_0-1 | <1$ or $| b_1+b_0-1 | <1$.

\noindent{\it Proof of Lemma~\ref{lem:error}.} From the law of total probability,  
\begin{eqnarray*}
 \pr(Q'=1 \mid S'=1)&=&(b_1+b_0-1) \pr(Q=1 \mid S'=1)+(1-b_0),\\
 \pr(Q'=1 \mid S'=0)&=&(b_1+b_0-1) \pr(Q=1 \mid S'=0)+(1-b_0),
\end{eqnarray*}
which imply
\begin{eqnarray}
\label{eqn:1} &&\pr(Q=1 \mid S'=1) =\frac{\pr(Q'=1 \mid S'=1)-(1-b_0)}{b_1+b_0-1},\\
\label{eqn:2}  &&\pr(Q=1 \mid S'=0) =\frac{\pr(Q'=1 \mid S'=0)-(1-b_0)}{b_1+b_0-1}.
\end{eqnarray}
Again, from  the law of total probability,  
\begin{eqnarray*}
&&\pr(Q=1\mid S'=1) =\pr(Q=1\mid S=1) a_1+\pr(Q=1\mid S=0) (1-a_1),\\
&&\pr(Q=1\mid S'=0) =\pr(Q=1\mid S=1) (1-a_0)+\pr(Q=1\mid S=0) a_0.
\end{eqnarray*}
Solving the above two equations, we have
\begin{eqnarray}
\label{eqn:3}&& \pr(Q=1 \mid S=1) =\frac{a_0 \pr(Q=1\mid S'=1)-(1-a_1)\pr(Q=1 \mid S'=0)}{a_1+a_0-1},\\
\label{eqn:4}&& \pr(Q=1 \mid S=0) =\frac{a_1 \pr(Q=1\mid S'=0)-(1-a_0)\pr(Q=1 \mid S'=1)}{a_1+a_0-1}.
\end{eqnarray}
Substituting (\ref{eqn:1}) and (\ref{eqn:2}) into (\ref{eqn:3}) and (\ref{eqn:4}), we obtain
\begin{eqnarray*}
\pr(Q=1 \mid S=1) &=&\frac{a_0\pr(Q'=1\mid S'=1)-(1-a_1)\pr(Q'=1\mid S'=0)}{(a_1+a_0-1)(b_1+b_0-1)}
-\frac{1-b_0}{b_1+b_0-1},\\
\pr(Q=1 \mid S=0) &=&\frac{a_1\pr(Q'=1\mid S'=0)-(1-a_0)\pr(Q'=1\mid S'=1)}{(a_1+a_0-1)(b_1+b_0-1)}
-\frac{1-b_0}{b_1+b_0-1},
\end{eqnarray*}
and
$$
\textsc{RD}_{Q'\mid S'} 
= (a_1+a_0-1)(b_1+b_0-1)\textsc{RD}_{Q\mid S}.
$$
\QEDB

\vspace{5mm}
\noindent{\it Proof of Theorem~\ref{thm:cace}.}  From Lemma~\ref{lem:error},
\begin{eqnarray*}
&& \CACE ' = \frac{\textsc{RD}_{Y'\mid Z'}}{\textsc{RD}_{D'\mid Z'}}=\frac{(\textsc{SN}_Y+\textsc{SP}_Y-1)(\textsc{SN}'_Z+\textsc{SP}'_Z-1)\textsc{RD}_{Y\mid Z}}{(\textsc{SN}_D+\textsc{SP}_D-1)(\textsc{SN}'_Z+\textsc{SP}'_Z-1)\textsc{RD}_{D\mid Z}}=\CACE  \times \frac{r_Y}{r_D}.
\end{eqnarray*}
\QEDB

\section{Bounds on $\CACE$ under non-differential measurement errors}\label{sec::boundsCACE}

\subsection{Proofs}

\noindent{\it Proof of Lemma~\ref{lem:bound}.} 
It is straightforward to write $\{ \pr(Z' =z', D'=d',Y'=y')  :   z',d',y'=0,1 \}$ in terms of the set $\{\pr(Z=z), \pr(U=u),\pr(Y_z=1\mid U=u) : z=0,1; u=a,n,c\}$ given $(\textsc{SN}'_Z,\textsc{SP}'_Z,\textsc{SN}_D,\textsc{SP}_D,\textsc{SN}_Y,\textsc{SP}_Y)$.

We then only need to show that we can express $\{\pr(Z=z), \pr(U=u),\pr(Y_z=1\mid U=u) : z=0,1; u=a,n,c\}$ in terms of $\{ \pr(Z' =z', D'=d',Y'=y')  :   z',d',y'=0,1 \}$ given $(\textsc{SN}'_Z,\textsc{SP}'_Z,\textsc{SN}_D,\textsc{SP}_D,\textsc{SN}_Y,\textsc{SP}_Y)$.

First, we can express $\pr(Z=1)$ as  
\begin{equation}
\label{eqn::pa} \pr(Z=1)=\textsc{SN}'_Z \pr(Z'=1)+(1-\textsc{SP}'_Z) \pr(Z'=0).
\end{equation}

Second, we express $\{ \pr(U=u): z=0,1; u=a,n,c\}$ in terms of $\{ \pr(Z' =z', D'=d',Y'=y')  :   z',d',y'=0,1 \}$ and $(\textsc{SN}'_Z,\textsc{SP}'_Z,\textsc{SN}_D,\textsc{SP}_D,\textsc{SN}_Y,\textsc{SP}_Y)$. From Assumption~\ref{asm:iv},
\begin{eqnarray*}
&&\pr(U=a)=\pr(D=1 \mid Z=0), \\
&&\pr(U=n)=\pr(D=0 \mid Z=1), \\
&&\pr(U=c)=1-\pr(U=a)-\pr(U=n).
\end{eqnarray*}
Then from Lemma~\ref{lem:error}, we can further express $\{ \pr(U=u):  u=a,n,c\}$ as 
\begin{eqnarray}
\label{eqn:ps:a}  
\pr(U=a)
&=&\frac{\textsc{SN}'_Z  \pr(D'=1 \mid Z'=0) -(1-\textsc{SP}'_Z)  \pr(D'=1 \mid Z'=1) }{(\textsc{SN}'_Z+\textsc{SP}'_Z-1)(\textsc{SN}_D+\textsc{SP}_D-1)} \nonumber \\
&& -\frac{1-\textsc{SP}_D}{\textsc{SN}_D+\textsc{SP}_D-1},\\
\label{eqn:ps:c} \pr(U=c)&=&\frac{\pr(D'=1 \mid Z'=1) -\pr(D'=1 \mid Z'=0) }{(\textsc{SN}'_Z+\textsc{SP}'_Z-1)(\textsc{SN}_D+\textsc{SP}_D-1)},\\
\label{eqn:ps:n} \pr(U=n)
&=&\frac{\textsc{SN}_D}{\textsc{SN}_D+\textsc{SP}_D-1} \nonumber \\
&& -\frac{\textsc{SP}'_Z  \pr(D'=1 \mid Z'=1)-(1-\textsc{SN}'_Z)  \pr(D'=1 \mid Z'=0) }{(\textsc{SN}'_Z+\textsc{SP}'_Z-1)(\textsc{SN}_D+\textsc{SP}_D-1)}.
\end{eqnarray}

Third, we express $\{\pr(Y_z=1\mid U=u) : z=0,1; u=a,n,c\}$ in terms of $\{ \pr(Z' =z', D'=d',Y'=y')  :   z',d',y'=0,1 \}$ and $(\textsc{SN}'_Z,\textsc{SP}'_Z,\textsc{SN}_D,\textsc{SP}_D,\textsc{SN}_Y,\textsc{SP}_Y)$.
By the law of total probability,
we  decompose the observed probabilities as
\begin{eqnarray}
\nonumber \pr(Y=1, D'=1\mid Z=1)
&=&\pr(Y_1=1 \mid U=a)  \pr(U=a) \textsc{SN}_D \\
\nonumber &&+ \pr(Y_1=1 \mid U=c)  \pr(U=c)  \textsc{SN}_D\\
\label{eqn::observed1} &&+\pr(Y_1=1 \mid U=n)   \pr(U=n)  (1-\textsc{SP}_D),\\
\nonumber  \pr(Y=1 \mid Z=1)
&=&\pr(Y_1=1 \mid U=a)   \pr(U=a)+ \pr(Y_1=1 \mid U=c)   \pr(U=c)\\
\label{eqn::observed2} &&+\pr(Y_1=1 \mid U=n)   \pr(U=n),\\
\nonumber \pr(Y=1, D'=1\mid Z=0)
&=&\pr(Y_0=1 \mid U=a)  \pr(U=a) \textsc{SN}_D \\
\nonumber &&+ \pr(Y_0=1 \mid U=c)  \pr(U=c)  (1-\textsc{SP}_D)\\
\label{eqn::observed3} &&+\pr(Y_0=1 \mid U=n)   \pr(U=n)  (1-\textsc{SP}_D),\\
\nonumber  \pr(Y=1 \mid Z=0)
&=&\pr(Y_0=1 \mid U=a)   \pr(U=a)+ \pr(Y_0=1 \mid U=c)   \pr(U=c)\\
\label{eqn::observed4} &&+\pr(Y_0=1 \mid U=n)   \pr(U=n).
\end{eqnarray}
Substituting ~\eqref{eqn:ps:a}--\eqref{eqn:ps:n} into ~\eqref{eqn::observed1}--\eqref{eqn::observed4}, we can obtain four equations for   $\{\pr(Y_z=1\mid U=u) : z=0,1; u=a,n,c\}$. 
From Assumption~\ref{asm:iv}(c), we can obtain two additional equations  $\pr(Y_1=1 \mid U=a) =\pr(Y_0=1 \mid U=a) $ and $\pr(Y_1=1 \mid U=n)=\pr(Y_0=1 \mid U=n)$. Solving them, we have 
\begin{eqnarray}
\nonumber \pr (Y_1=1 \mid U=n)&=&\pr (Y_0=1 \mid U=n)\\
\label{eqn:Y1Un} &=&\frac{\textsc{SN}_D   \pr(Y=1\mid Z=1)-\pr(Y=1, D'=1\mid Z=1)}{\textsc{SN}_D-\pr(D'=1 \mid Z=1)},\\
\nonumber\pr (Y_1=1 \mid U=a)&=&\pr (Y_0=1 \mid U=a)\\
\label{eqn:Y1Ua}\label{eqn}&=& \frac{\pr(Y=1, D'=1\mid Z=0)-(1-\textsc{SP}_D)   \pr(Y=1\mid Z=0)}{\pr(D'=1 \mid Z=0)-(1-\textsc{SP}_D)},\\
\label{eqn:Y1Uc}\pr (Y_1=1 \mid U=c)&=& \frac{\textsc{RD}_{YD'\mid Z}}{\textsc{RD}_{D'\mid Z}} 
-(1-\textsc{SP}_D)\times \frac{\textsc{RD}_{Y\mid Z}}{\textsc{RD}_{D'\mid Z}},\\
\label{eqn:Y0Uc}\pr (Y_0=1 \mid U=c)&=& \frac{\textsc{RD}_{YD'\mid Z}}{\textsc{RD}_{D'\mid Z}} 
-\textsc{SN}_D\times  \frac{\textsc{RD}_{Y\mid Z}}{\textsc{RD}_{D'\mid Z}}.
\end{eqnarray}

We use Lemma~\ref{lem:error} to obtain
\begin{eqnarray}
\nonumber &&\pr(Y=1,D'=1\mid Z=1) \\
\nonumber&=&\frac{\textsc{SP}'_Z \pr(Y'=1,D'=1 \mid Z'=1)-(1-\textsc{SN}'_Z)  \pr(Y'=1,D'=1 \mid Z'=0)  }{(\textsc{SN}'_Z+\textsc{SP}'_Z-1)(\textsc{SN}_Y+\textsc{SP}_Y-1)}\\
\nonumber&&-\frac{\textsc{SP}'_Z(1-\textsc{SP}_Y)\pr(D'=1\mid Z'=1)}{(\textsc{SN}'_Z+\textsc{SP}'_Z-1)(\textsc{SN}_Y+\textsc{SP}_Y-1)}+\frac{(1-\textsc{SN}'_Z)(1-\textsc{SP}_Y)\pr(D'=1\mid Z'=0)}{(\textsc{SN}'_Z+\textsc{SP}'_Z-1)(\textsc{SN}_Y+\textsc{SP}_Y-1)}\\
\label{eqn:Y1D1Z1}
\end{eqnarray}
and
\begin{eqnarray}
\nonumber && \pr(Y=1,D'=1\mid Z=0) \\
\nonumber&=&\frac{\textsc{SN}'_Z  \pr(Y'=1,D'=1 \mid Z'=0) -(1-\textsc{SP}'_Z)  \pr(Y'=1,D'=1 \mid Z'=1) }{(\textsc{SN}'_Z+\textsc{SP}'_Z-1)(\textsc{SN}_Y+\textsc{SP}_Y-1)}\\
\nonumber&&-\frac{\textsc{SN}'_Z(1-\textsc{SP}_Y)\pr(D'=1\mid Z'=0)}{(\textsc{SN}'_Z+\textsc{SP}'_Z-1)(\textsc{SN}_Y+\textsc{SP}_Y-1)}+\frac{(1-\textsc{SP}'_Z)(1-\textsc{SP}_Y)\pr(D'=1\mid Z'=1)}{(\textsc{SN}'_Z+\textsc{SP}'_Z-1)(\textsc{SN}_Y+\textsc{SP}_Y-1)}.\\
\label{eqn:Y1D1Z0}&&
\end{eqnarray}

Substituting~\eqref{eqn:Y1D1Z1} and~\eqref{eqn:Y1D1Z0}  into~\eqref{eqn:Y1Un}--\eqref{eqn:Y0Uc}, we have
{\small 
\begin{eqnarray}
\nonumber&&\pr (Y_1=1 \mid U=n)=\pr (Y_0=1 \mid U=n)\\
\nonumber&=& \frac{\textsc{SN}_D}{\textsc{SN}_Y+\textsc{SP}_Y-1} \times \frac{\textsc{SP}'_Z  \pr(Y'=1\mid Z'=1)-(1-\textsc{SN}'_Z)  \pr(Y'=1\mid Z'=0)}{(\textsc{SN}'_Z+\textsc{SP}'_Z-1)\textsc{SN}_D-\textsc{SP}'_Z \pr(D'=1 \mid Z'=1)+(1-\textsc{SN}'_Z)\pr(D'=1\mid Z'=0)}\\
\nonumber&&-\frac{\textsc{SP}'_Z  \pr(Y'=1,D'=1\mid Z'=1)-(1-\textsc{SN}'_Z)  \pr(Y'=1,D'=1\mid Z'=0)}{(\textsc{SN}_Y+\textsc{SP}_Y-1)\{(\textsc{SN}'_Z+\textsc{SP}'_Z-1)\textsc{SN}_D-\textsc{SP}'_Z \pr(D'=1 \mid Z'=1)+(1-\textsc{SN}'_Z)\pr(D'=1\mid Z'=0)\}}\\
\label{eqn:py:n} &&-\frac{1-\textsc{SP}_Y}{\textsc{SN}_Y+\textsc{SP}_Y-1},
\end{eqnarray}
\begin{eqnarray}
\nonumber&&\pr (Y_1=1 \mid U=a)=\pr (Y_0=1 \mid U=a)\\
\nonumber&=&\frac{\textsc{SN}'_Z  \pr(Y'=1,D'=1 \mid Z'=0) -(1-\textsc{SP}'_Z) \pr(Y'=1,D'=1 \mid Z'=1) }{(\textsc{SN}_Y+\textsc{SP}_Y-1)\{\textsc{SN}'_Z \pr(D'=1 \mid Z'=0)-(1-\textsc{SP}'_Z)\pr(D'=1\mid Z'=1)-(\textsc{SN}'_Z+\textsc{SP}'_Z-1)(1-\textsc{SP}_D) \}}\\
\nonumber&&-\frac{1-\textsc{SP}_D}{\textsc{SN}_Y+\textsc{SP}_Y-1}\times \frac{\textsc{SN}'_Z  \pr(Y'=1\mid Z'=0) -(1-\textsc{SP}'_Z)  \pr(Y'=1 \mid Z'=1)}{\textsc{SN}'_Z \pr(D'=1 \mid Z'=0)-(1-\textsc{SP}'_Z)\pr(D'=1\mid Z'=1)-(\textsc{SN}'_Z+\textsc{SP}'_Z-1)(1-\textsc{SP}_D)}\\
\label{eqn:py:a}&&-\frac{1-\textsc{SP}_Y}{\textsc{SN}_Y+\textsc{SP}_Y-1},\\
\label{eqn:py:c1} &&\pr (Y_1=1 \mid U=c)=\frac{1}{\textsc{SN}_Y+\textsc{SP}_Y-1}\times \left \{ \frac{\textsc{RD}_{Y'D'\mid Z'}}{\textsc{RD}_{D'\mid Z'}} 
-(1-\textsc{SP}_D) \times \frac{\textsc{RD}_{Y'\mid Z'}}{\textsc{RD}_{D'\mid Z'}} -(1-\textsc{SP}_Y) \right \},\\
 \label{eqn:py:c0} &&\pr (Y_0=1 \mid U=c)=\frac{1}{\textsc{SN}_Y+\textsc{SP}_Y-1}\times \left \{ \frac{\textsc{RD}_{Y'D'\mid Z'}}{\textsc{RD}_{D'\mid Z'}} 
-\textsc{SN}_D \times \frac{\textsc{RD}_{Y'\mid Z'}}{\textsc{RD}_{D'\mid Z'}} -(1-\textsc{SP}_Y)\right \}.
\end{eqnarray}
}

From \eqref{eqn::pa}, \eqref{eqn:ps:a}--\eqref{eqn:ps:n} and \eqref{eqn:py:n}--\eqref{eqn:py:c0},  we can express $\{\pr(Z=z), \pr(U=u),\pr(Y_z=1\mid U=u) : z=0,1; u=a,n,c\}$ in terms of $\{ \pr(Z' =z', D'=d',Y'=y')  :   z',d',y'=0,1 \}$ and $(\textsc{SN}'_Z,\textsc{SP}'_Z,\textsc{SN}_D,\textsc{SP}_D,\textsc{SN}_Y,\textsc{SP}_Y)$. 
\QEDB

\vspace{5mm}
From Lemma~\ref{lem:bound}, if we know the sensitivities and specificities, then we can recover the joint distribution of all the potential outcomes. Furthermore, the conditions 
\begin{eqnarray}
\label{eq::solvingconditions}
\{ 0\leq  \pr(Z=z) \leq 1, 0\leq \pr(U=u) \leq 1, 0\leq \pr(Y_z=1\mid U=u) \leq 1 : z=0,1; u= a,n,c \} \nonumber \\
\end{eqnarray}
induce sharp bounds on the sensitivities and specificities, which in turn induce sharp bounds on $\CACE$.

\vspace{5mm}
\noindent {\it Proof of Theorem~\ref{thm:bound:Y}.} If only $Y$ is mismeasured, $\textsc{SP}_D=\textsc{SN}_D=\textsc{SP}'_Z=\textsc{SN}'_Z=1$. In this case, the formulas of $\{ \pr(Z=z), \pr(U=u): z=0,1; u=a,n,c\}$ in \eqref{eqn::pa} and \eqref{eqn:ps:a}--\eqref{eqn:ps:n} do not depend on $(\textsc{SN}_Y, \textsc{SP}_Y)$, and thus do not
 provide any information for them. We then consider only the inequalities
 \begin{eqnarray}
\{  0\leq \pr(Y_z=1  \mid U=u)\leq 1: z=0,1; u=a,n,c \}. \label{eq::misY-ineq}
 \end{eqnarray}
 
From (\ref{eqn:py:n})--(\ref{eqn:py:c0}),   
{\small
\begin{eqnarray}
\label{eqn:th2:y1} &&\pr (Y_1=1 \mid U=n)=\pr (Y_0=1 \mid U=n)
= \frac{\pr(Y'=1\mid D=0,Z=1)-(1-\textsc{SP}_Y)}{\textsc{SN}_Y+\textsc{SP}_Y-1},\\
\label{eqn:th2:y2}  &&\pr (Y_1=1 \mid U=a)=\pr (Y_0=1 \mid U=a)
= \frac{\pr(Y'=1\mid D=1,Z=0)-(1-\textsc{SP}_Y)}{\textsc{SN}_Y+\textsc{SP}_Y-1},\\
\label{eqn:th2:y3}  &&\pr(Y_1=1  \mid U=c)
= \frac{1}{\textsc{SN}_Y+\textsc{SP}_Y-1}\times \frac{\textsc{RD}_{Y'D\mid Z}}{\textsc{RD}_{D\mid Z}} 
-\frac{1-\textsc{SP}_Y}{\textsc{SP}_Y+\textsc{SN}_Y-1},\\
\label{eqn:th2:y4}  &&\pr(Y_0=1  \mid U=c)
= \frac{1}{\textsc{SN}_Y+\textsc{SP}_Y-1}\times \frac{\textsc{RD}_{Y'(1-D)\mid (1-Z)}}{\textsc{RD}_{D\mid Z}} 
-\frac{1-\textsc{SP}_Y}{\textsc{SP}_Y+\textsc{SN}_Y-1}.
\end{eqnarray} 
}
Solving \eqref{eq::misY-ineq}, we obtain
{\small
\begin{eqnarray}
\textsc{SN}_Y &\geq& \max \left \{\pr(Y'=1 \mid D=0,  Z=1),\pr(Y'=1 \mid D=1, Z=0),  
 \frac{\textsc{RD}_{Y'D\mid Z}}{\textsc{RD}_{D\mid Z}}, \frac{\textsc{RD}_{Y'(1-D)\mid (1-Z)}}{\textsc{RD}_{D\mid Z}}\right\}, \nonumber\\
  \label{eq::sny}\\
\textsc{SP}_Y &\geq& 1-\min \left \{\pr(Y'=1 \mid D=0, Z=1),\pr(Y'=1 \mid D=1, Z=0),  
  \frac{\textsc{RD}_{Y'D\mid Z}}{\textsc{RD}_{D\mid Z}}, \frac{\textsc{RD}_{Y'(1-D)\mid (1-Z)}}{\textsc{RD}_{D\mid Z}}\right\}. \nonumber \\
  \label{eq::spy}
\end{eqnarray} 
}
After rearrangement, only one of $\textsc{SN}_Y$ and $\textsc{SP}_Y$ appears in each of the inequalities in \eqref{eq::misY-ineq}. As a result, 
the bounds on $\textsc{SN}_Y$ and $\textsc{SP}_Y$ are both attainable. Thus, we can then obtain the sharp bounds on $r_Y$ by summing \eqref{eq::sny} and \eqref{eq::spy}.
Then,
$\CACE ' \leq \CACE  \leq \CACE '/(M_Y- N_Y)$.
\QEDB

\vspace{5mm}
\noindent {\it Proof of Corollary~\ref{cor:testY}.}  For $\textsc{SN}_Y$ and $\textsc{SP}_Y $, 
 the lower bounds must be smaller than or equal to $1$. Under $\CACE ' \geq 0$, these require
\begin{eqnarray*}
&& 0\leq \frac{\textsc{RD}_{Y'D\mid Z}}{\textsc{RD}_{D\mid Z}}\leq 1, \quad 0 \leq \frac{\textsc{RD}_{Y'(1-D)\mid (1-Z)}}{\textsc{RD}_{D\mid Z}} \leq 1,
\end{eqnarray*}
which are equivalent to the inequalities in Corollary~\ref{cor:testY}. Under $\CACE ' < 0$, we can obtain the same conditions.
\QEDB

\vspace{5mm}
\noindent {\it Proof of Theorem~\ref{thm:bound:D}.} If only $D$ is mismeasured, $\textsc{SP}_Y=\textsc{SN}_Y=\textsc{SP}'_Z=\textsc{SN}'_Z=1$.
In this case, the formula of $\pr(Z=1)$ does not depend on $(\textsc{SN}_D, \textsc{SP}_D)$, and thus does not
 provide any information for them. We then consider only the inequalities
 \begin{eqnarray}
\{ 0\leq  \pr(U=u) \leq 1, 0\leq \pr(Y_z=1  \mid U=u)\leq 1: z=0,1; u=a,n,c \}. \label{eq::misD-ineq}
 \end{eqnarray}
 
 From (\ref{eqn:ps:a})--(\ref{eqn:ps:n}),   
\begin{eqnarray*}
&&\pr(U=a)=\frac{\pr(D'=1 \mid Z=0)-(1-\textsc{SP}_D)}{\textsc{SN}_D+\textsc{SP}_D-1},\\
&&\pr(U=n)=\frac{\textsc{SN}_D-\pr(D'=1 \mid Z=1)}{\textsc{SN}_D+\textsc{SP}_D-1},\\
&&\pr(U=c)=\frac{\pr(D'=1 \mid Z=1)-\pr(D'=1 \mid Z=0)}{\textsc{SN}_D+\textsc{SP}_D-1}.
\end{eqnarray*}
From (\ref{eqn:py:n})--(\ref{eqn:py:c0}),  
\begin{eqnarray*}
\pr (Y_1=1 \mid U=n) &=& \pr (Y_0=1 \mid U=n) \\
&=&\frac{\textsc{SN}_D  \times  \pr(Y=1\mid Z=1)-\pr(Y=1, D'=1\mid Z=1)}{\textsc{SN}_D-\pr(D'=1 \mid Z=1)},\\
 \pr (Y_1=1 \mid U=a)&=&\pr (Y_0=1 \mid U=a) \\
&=& \frac{\pr(Y=1, D'=1\mid Z=0)-(1-\textsc{SP}_D)   \times \pr(Y=1\mid Z=0)}{\pr(D'=1 \mid Z=0)-(1-\textsc{SP}_D)},\\
\pr (Y_1=1 \mid U=c) &=& \frac{\textsc{RD}_{YD'\mid Z}}{\textsc{RD}_{D'\mid Z}}
-(1-\textsc{SP}_D) \times \frac{\textsc{RD}_{Y\mid Z}}{\textsc{RD}_{D'\mid Z}},\\
\pr (Y_0=1 \mid U=c) &=& \frac{\textsc{RD}_{YD'\mid Z}}{\textsc{RD}_{D'\mid Z}} -\textsc{SN}_D \times  \frac{\textsc{RD}_{Y\mid Z}}{\textsc{RD}_{D'\mid Z}} .
\end{eqnarray*}
Solving \eqref{eq::misD-ineq}, we can obtain the bounds on $\textsc{SN}_D$ and $\textsc{SP}_D$. When $\CACE ' \geq 0$, we have 
\begin{eqnarray}
&& \max_{z,y=0,1}\left\{\pr(D'=1\mid Z=z),\pr(D'=1 \mid Y=y,Z=1),\frac{\textsc{RD}_{(1-Y)D'\mid (1-Z)}}{\textsc{RD}_{Y\mid Z}}\right\}  \nonumber \\
&\leq& \textsc{SN}_D \leq \min\left \{1,\frac{\textsc{RD}_{YD'\mid Z}}{\textsc{RD}_{Y\mid Z}}\right\} , \label{eq::snd} \\ 
&& 1-\min_{z,y=0,1}\left\{\pr(D'=1\mid Z=z),\pr(D'=1 \mid Y=y,Z=0),\frac{\textsc{RD}_{YD'\mid Z}}{\textsc{RD}_{Y\mid Z}}\right\} \nonumber\\
&\leq& \textsc{SP}_D \leq 1-\max\left \{0,\frac{\textsc{RD}_{(1-Y)D'\mid (1-Z)}}{\textsc{RD}_{Y\mid Z}}\right\}. \label{eq::spd}
\end{eqnarray}
Because only one of $\textsc{SN}_D$ and $\textsc{SP}_D$ appears  in each of the inequalities in \eqref{eq::misD-ineq} after rearrangement, 
the bounds on $\textsc{SN}_D$ and $\textsc{SP}_D$ are both attainable. We can then obtain the sharp bounds on $r_D$  by summing \eqref{eq::snd} and \eqref{eq::spd}. 
We can then obtain the sharp bounds on $\CACE$.
\QEDB

\vspace{5mm}
Before proving Corollary~\ref{cor:testD}, we give a simple lemma.

\begin{lemma}
\label{lemma::mono-measurementerrorD}
Under Assumption \ref{asm:iv}(b), if only $D$ is mismeasured with $r_D > 0$, then $\textsc{RD}_{D'|Z} \geq 0.$
\end{lemma}

\noindent {\it Proof of Lemma~\ref{lemma::mono-measurementerrorD}.} 
Assumption \ref{asm:iv}(b) implies that $ \textsc{RD}_{D|Z} \geq 0 .$
Using Lemma \ref{lem:error}, we have $\textsc{RD}_{D'|Z} = \textsc{RD}_{D|Z} \times r_D \geq 0.$
\QEDB

\vspace{5mm}
\noindent {\it Proof of Corollary~\ref{cor:testD}.}  For $\textsc{SN}_D$ and $\textsc{SP}_D$, the lower bounds must be smaller than or equal to 1 and the upper bounds must be larger than or equal to 0. Moreover,
the lower bounds must  be smaller than or equal to their upper bounds. These require
\begin{eqnarray}
\label{eqn::testableMisD1} &&  \frac{\textsc{RD}_{YD'\mid Z}}{\textsc{RD}_{Y\mid Z}} \geq 0, \quad  \frac{\textsc{RD}_{(1-Y)D'\mid (1-Z)}}{\textsc{RD}_{Y\mid Z}} \leq 1,\\
\label{eqn::testableMisD2}&&  \pr(D'=1 \mid Y=y,Z=1) \leq  \frac{\textsc{RD}_{YD'\mid Z}}{\textsc{RD}_{Y\mid Z}},  \hspace{1.6cm} (y=0,1),\\
\label{eqn::testableMisD3} &&   \pr(D'=1 \mid Y=y,Z=0) \geq  \frac{\textsc{RD}_{(1-Y)D'\mid (1-Z)}}{\textsc{RD}_{Y\mid Z}}, \quad (y=0,1), \\
\label{eqn::redundant1} &&  \frac{\textsc{RD}_{(1-Y)D'\mid (1-Z)}}{\textsc{RD}_{Y\mid Z}} \leq \frac{\textsc{RD}_{YD'\mid Z}}{\textsc{RD}_{Y\mid Z}} ,\\
\label{eqn::redundant2} && \pr(D'=1\mid Z=z ) \geq     \frac{\textsc{RD}_{(1-Y)D'\mid (1-Z)}}{\textsc{RD}_{Y\mid Z}}, \quad \hspace{1.2cm} (z=0,1),\\
\label{eqn::redundant3} && \pr(D'=1\mid Z=z )  \leq \frac{\textsc{RD}_{YD'\mid Z}}{\textsc{RD}_{Y\mid Z}}, \quad  \hspace{2.45cm} (z=0,1).
\end{eqnarray}
When $\CACE ' \geq 0$, Lemma \ref{lemma::mono-measurementerrorD} ensures \eqref{eqn::redundant1}. Moreover, \eqref{eqn::testableMisD3} implies \eqref{eqn::redundant2} with $z=0$, and \eqref{eqn::testableMisD2} implies \eqref{eqn::redundant3} with $z=1$. Lemma \ref{lemma::mono-measurementerrorD} further ensures \eqref{eqn::redundant2} with $z=1$ and \eqref{eqn::redundant3} with $z=0$. Therefore, \eqref{eqn::redundant1} to \eqref{eqn::redundant3} are redundant. The remaining conditions \eqref{eqn::testableMisD1}--\eqref{eqn::testableMisD3} 
are equivalent to the inequalities in Corollary~\ref{cor:testD}.
\QEDB

\vspace{5mm}
Next, we show that the conditions in Corollary~\ref{cor:testD} are weaker than the following conditions in \citet{balke1997bounds}: 
\begin{eqnarray}
\pr(Y=y,D'=1 \mid Z=1) &\geq& \pr(Y=y, D'=1 \mid Z=0), \quad (y=0,1), \label{eqn::pearl1}\\
\pr(Y=y,D'=0 \mid Z=0) &\geq& \pr(Y=y, D'=0 \mid Z=1), \quad (y=0,1).\label{eqn::pearl2}
\end{eqnarray}
These are the testable conditions for the binary instrumental variable model without measurement errors.

\noindent {\it Proof. }First, \eqref{eqn::testableMisD1} is equivalent to ~\eqref{eqn::pearl1} with $y=1$ and ~\eqref{eqn::pearl2} with $y=0$.

Second, we show ~\eqref{eqn::pearl1} and ~\eqref{eqn::pearl2} imply \eqref{eqn::testableMisD2}. From ~\eqref{eqn::pearl1} and ~\eqref{eqn::pearl2} with $y=1$, we have 
\begin{eqnarray*}
\frac{\pr(Y=1,D'=1\mid Z=1)}{ \sum_{d'=0,1}\pr(Y=1,D'=d'\mid Z=1)} \geq \frac{\pr(Y=1,D'=1\mid Z=0)}{\sum_{d'=0,1}\pr(Y=1,D'=d'\mid Z=0)}.
\end{eqnarray*}
Therefore,  $\pr(D'=1\mid Y=1, Z=1)\geq \pr(D'=1\mid Y=1, Z=0)$, which is equivalent to ~\eqref{eqn::testableMisD2} with $y=1$.
From $\textsc{RD}_{Y\mid Z} \geq 0$, we have 
\begin{eqnarray*}
&&\pr(Y=0\mid Z=0)\pr(D'=1\mid Y=0,Z=1)\\
&=&\pr(Y=0\mid Z=0)\{1-\pr(D'=0\mid Y=0,Z=1)\}\\
&=&\pr(Y=0\mid Z=0)-\pr(Y=0\mid Z=0)\pr(D'=0\mid Y=0,Z=1)\\
&\leq&\pr(Y=0\mid Z=0)-\pr(Y=0\mid Z=1)\pr(D'=0\mid Y=0,Z=1)\\
&=&\pr(Y=0,D'=1\mid Z=0).
\end{eqnarray*}
Therefore,
\begin{eqnarray*}
\pr(D'=1\mid Z=1) &\geq& \pr(D'=1\mid Z=0)\\
&=& \pr(Y=1,D'=1\mid Z=0)+\pr(Y=0,D'=1\mid Z=0)\\
&\geq &\pr(Y=1,D'=1\mid Z=0)+ \pr(Y=0\mid Z=0)\pr(D'=1\mid Y=0,Z=1),
\end{eqnarray*}
which is equivalent to ~\eqref{eqn::testableMisD2} with $y=0$.

Third, we show ~\eqref{eqn::pearl1} and ~\eqref{eqn::pearl2} imply \eqref{eqn::testableMisD3}.
From ~\eqref{eqn::pearl1} with $y=0$ and $\textsc{RD}_{Y\mid Z} \geq 0$,  we have 
\begin{eqnarray*}
\pr(D'=1\mid Z=0)
& =&  \pr(Y=1,D'=1\mid Z=0)+\pr(Y=0,D'=1\mid Z=0)\\
&\leq &\pr(Y=1\mid Z=0)\pr(D'=1\mid Y=1, Z=0)+\pr(Y=0,D'=1\mid Z=1)\\
&\leq&\pr(Y=1\mid Z=1)\pr(D'=1\mid Y=1, Z=0)+\pr(Y=0,D'=1\mid Z=1),
\end{eqnarray*}
which is equivalent to ~\eqref{eqn::testableMisD3} with $y=1$.
From ~\eqref{eqn::pearl1} with $y=0$ and $\textsc{RD}_{Y\mid Z} \geq 0$, we have 
\begin{eqnarray*}
\frac{\pr(Y=0,D'=1 \mid Z=1)}{\pr(Y=0 \mid Z=1)} \geq  \frac{\pr(Y=0,D'=1 \mid Z=0)}{\pr(Y=0 \mid Z=0)}.
\end{eqnarray*}
Therefore, $\pr(D'=1\mid Y=0,Z=1) \geq \pr(D'=1\mid Y=0,Z=0) $, which is equivalent to ~\eqref{eqn::testableMisD3} with $y=0$. \QEDB

\subsection{Bounds on $\CACE$ under simultaneous measurement errors}
\label{sec::numerical}
It is complicated to obtain closed-form bounds under simultaneous measurement errors of more than one elements of $(Z,D,Y)$. We provide a general strategy for calculating the sharp bounds numerically.

From Lemma~\ref{lem:bound}, we can express $\{ \pr(Z' =z', D'=d',Y'=y'):   z',d',y'=0,1 \}$ in terms of $\{\pr(Z=z), \pr(U=u),\pr(Y_z=1\mid U=u) : z=0,1; u=a,n,c\}$ and $(\textsc{SN}'_Z,\textsc{SP}'_Z,\textsc{SN}_D,\textsc{SP}_D,\textsc{SN}_Y,\textsc{SP}_Y)$. Therefore, we obtain $8-1 = 7$  equality constraints for $\{\pr(Z=z), \pr(U=u),\pr(Y_z=1\mid U=u) : z=0,1; u=a,n,c\}$ and $(\textsc{SN}'_Z,\textsc{SP}'_Z,\textsc{SN}_D,\textsc{SP}_D,\textsc{SN}_Y,\textsc{SP}_Y)$. 
Using linear or non-linear programming, we can numerically calculate the bounds by minimizing and maximizing $\CACE$ under the equality constraints and the inequality constraints~\eqref{eq::solvingconditions}.

\section{Results under strong monotonicity}\label{sec::strongmono}

\subsection{Proofs}
\noindent {\it Proof of Theorem~\ref{thm:bound:str:Y}.}  
If only $Y$ is mismeasured,   
\begin{eqnarray*}
&&\CACE ' =\frac{\pr(Y'=1 \mid Z=1)-\pr(Y'=1 \mid Z=0)}{\pr(D=1 \mid Z=1)}, \quad \CACE =\CACE '/r_Y.
\end{eqnarray*}
In this case, the formulas of $\{ \pr(Z=z), \pr(U=u): z=0,1; u=n,c\}$ do not depend on $(\textsc{SN}_Y, \textsc{SP}_Y)$, and thus do not provide any information for them. Therefore, we consider only the inequalities in \eqref{eq::misY-ineq} based on the following probabilities:
\begin{eqnarray}
&&\pr (Y_1=1 \mid U=n)=\pr (Y_0=1 \mid U=n)
= \frac{\pr(Y'=1\mid D=0,Z=1)-(1-\textsc{SP}_Y)}{\textsc{SN}_Y+\textsc{SP}_Y-1},  \label{eq::misymono1} \\
 &&\pr(Y_1=1  \mid U=c)
=  \frac{\pr(Y'=1\mid D=1,Z=1)-(1-\textsc{SP}_Y)}{\textsc{SN}_Y+\textsc{SP}_Y-1},  \label{eq::misymono2} \\
 &&\pr(Y_0=1  \mid U=c)
= \frac{\pr(Y'=1 \mid D=1,Z=1)-\CACE '}{\textsc{SN}_Y+\textsc{SP}_Y-1}
-\frac{1-\textsc{SP}_Y}{\textsc{SP}_Y+\textsc{SN}_Y-1}.\label{eq::misymono3}
\end{eqnarray} 
Using \eqref{eq::misymono1}--\eqref{eq::misymono3} to solve \eqref{eq::misY-ineq}, we obtain 
\begin{eqnarray}
&\text{SP}_Y &\geq 1- \min \{\pr(Y'=1\mid D=0,Z=1),\pr(Y'=1 \mid D=1,Z=1)-\CACE '\}, \label{eq::spy-mono}\\
&\text{SN}_Y &\geq  \max \{\pr(Y'=1\mid D=0,Z=1),\pr(Y'=1\mid D=1,Z=1)\}. \label{eq::sny-mono}
\end{eqnarray}
After rearrangement, only one of $\textsc{SN}_Y$ and $\textsc{SP}_Y$ appears in each of the inequalities in \eqref{eq::misY-ineq}. As a result, 
the bounds \eqref{eq::spy-mono} and \eqref{eq::sny-mono} are both attainable.  Thus, we obtain the sharp bounds on $r_Y$  by summing \eqref{eq::spy-mono} and \eqref{eq::sny-mono}, i.e., $r_Y \geq ( M_Y^{\textup{m}} - N_Y^{\textup{m}}  )$, where
\begin{eqnarray*}
M_Y^{\textup{m}} - N_Y^{\textup{m}}&=&  \max \{\pr(Y'=1\mid D=0,Z=1),\pr(Y'=1\mid D=1,Z=1)\}\\
&&- \min \{\pr(Y'=1\mid D=0,Z=1),\pr(Y'=1 \mid D=1,Z=1)-\CACE '\} .
\end{eqnarray*}
Therefore, 
$\CACE ' \leq \CACE  \leq \CACE '/( M_Y^{\textup{m}} - N_Y^{\textup{m}} )$. 
\QEDB

\vspace{5mm}
We give a more general version of Theorem~\ref{thm:bound:str:D} without the condition in \eqref{eq::conditionD}.

\setcounter{theorem}{0}
\renewcommand {\thetheorem} {S\arabic{theorem}}
\begin{theorem}
\label{thm:bound:str:D-general} 
Suppose that $\CACE ' \geq 0$ and  only $D$ is mismeasured.
Define 
\begin{eqnarray*}
S_D &=& 
\Big\{ \frac{\pr(Y=1,D'=1 \mid Z=1)-\textsc{RD}_{Y\mid Z} \times \max_{y=0,1}\{\pr(D'=1 \mid Y=y, Z=1)\}}{\pr(Y=1\mid Z=0)},\\
&&  \pr(D'=1 \mid Y=y, Z=1)   \Big\}.
\end{eqnarray*}
Under Assumptions~\ref{asm:iv} and~\ref{asm:str}, the  sharp upper bound on 
 $\CACE $ is 
\begin{eqnarray*}
\label{eqn::upperbound:str:D}
 \max \left\{ \pr(D'=1\mid Y=1,Z=1), \frac{\pr(Y=1,D'=1\mid Z=1)- \min S_{D}\times \pr(Y=1\mid Z=1) }{\pr(D'=1 \mid Z=1)- \min S_{D}} \right \},
 \end{eqnarray*}
 and   the sharp lower bound on $\CACE$ is
 \begin{eqnarray*}
 \nonumber &\min& \left \{  \frac{\max_{y=0,1}\pr(D'=1 \mid Y=y, Z=1)\times  \textsc{RD}_{Y\mid Z}}{\pr(D'=1 \mid Z=1)}, \right. \\
\label{eqn::lowerbound:str:D} &&  \left.  \frac{\{\max_{y=0,1}\pr(D'=1 \mid Y=y, Z=1)-\min S_{D}\} \times \textsc{RD}_{Y\mid Z}}{\pr(D'=1 \mid Z=1)-\min S_{D}} \right \}.
\end{eqnarray*}
\end{theorem}

\noindent {\it Proof of Theorem~\ref{thm:bound:str:D-general}.}  
First, $\pr(Z=z)$ does not depend on $(\textsc{SN}^1_D,\textsc{SP}^1_D)$, and thus the condition $0 \leq \pr(Z=1)\leq 1$
do not provide any information for them.
We need only to express $\{ \pr(U=u),\pr(Y_z=1\mid U=u) : z=0,1; u=n,c\}$ in terms of $\{ \pr(Z =z, D'=d',Y=y)  :   z,d',y=0,1 \}$ and $(\textsc{SN}^1_D,\textsc{SP}^1_D)$. 
The proportions of principal strata are
\begin{eqnarray*}
\pr(U=c)=\frac{\pr(D'=1\mid Z=1)-(1-\textsc{SP}_D^1)}{\textsc{SN}_D^1+\textsc{SP}_D^1-1}, \quad
\pr(U=n)=\frac{\textsc{SN}_D^1-\pr(D'=1\mid Z=1)}{\textsc{SN}_D^1+\textsc{SP}_D^1-1}.
\end{eqnarray*}
We decompose the observed probabilities into 
\begin{eqnarray*}
\pr(Y=1,D'=1\mid Z=1) &=&\pr(Y_1=1\mid U=c)\pr(U=c) \textsc{SN}_D^1\\
&&+\pr(Y_1=1\mid U=n)\pr(U=n) (1-\textsc{SP}_D^1),\\
\pr(Y=1\mid Z=1) &=&\pr(Y_1=1\mid U=c)\pr(U=c) +\pr(Y_1=1\mid U=n)\pr(U=n).
\end{eqnarray*}
Solving the above two equations, we have   
\begin{eqnarray*}
\pr(Y_1=1\mid U=c)  &=& \frac{\pr(Y=1,D'=1\mid Z=1)-(1-\textsc{SP}_D^1)\times  \pr(Y=1\mid Z=1) }{\pr(D'=1\mid Z=1)-(1-\textsc{SP}_D^1)},\\
\pr(Y_0=1\mid U=n) &=& \pr(Y_1=1\mid U=n) \\
&=&\frac{ \textsc{SN}_D^1 \times \pr(Y=1\mid Z=1) -\pr(Y=1,D'=1\mid Z=1) }{\textsc{SN}_D^1-\pr(D'=1\mid Z=1)}.
\end{eqnarray*}
From  the following decomposition of the outcome distribution in the control group
\begin{eqnarray*}
&&\pr(Y=1\mid Z=0) =\pr(Y_0=1\mid U=c)\pr(U=c) +\pr(Y_0=1\mid U=n)\pr(U=n),
\end{eqnarray*}
we obtain
\begin{eqnarray*}
\pr(Y_0=1\mid U=c) &=& \frac{\pr(Y=1,D'=1\mid Z=1)-(1-\textsc{SP}_D^1)\pr(Y=1\mid Z=1) }{\pr(D'=1 \mid Z=1)-(1-\textsc{SP}_D^1)}\\
&&-\frac{(\textsc{SN}_D^1+\textsc{SP}_D^1-1)\textsc{RD}_{Y\mid Z}}{\pr(D'=1 \mid Z=1)-(1-\textsc{SP}_D^1)}.\\
\end{eqnarray*}

Second, we derive the bounds on $\textsc{SN}_D^1$ and $\textsc{SP}_D^1$ by solving the inequalities in 
 \begin{eqnarray*}
\{ 0\leq  \pr(U=u) \leq 1, 0\leq \pr(Y_z=1  \mid U=u)\leq 1: z=0,1; u=c,n \}. 
 \end{eqnarray*}
Solving $\{  0 \leq \pr(U=u) \leq 1: u = c, n\} $, we have
\begin{eqnarray}
\textsc{SN}_D^1 \geq \pr(D'=1 \mid Z=1), \quad \textsc{SP}_D^1 \geq 1-\pr(D'=1 \mid Z=1). \label{eqn::strD-snsp1}
\end{eqnarray}
Solving $0 \leq \pr(Y_1=1 \mid U=c) \leq 1$ and $0 \leq \pr(Y_1=1 \mid U=n) \leq 1$, we have
\begin{eqnarray}
\label{eqn::strD-sn2} \textsc{SN}_D^1 &\geq& \pr(D'=1 \mid Y=y, Z=1), \quad \hspace{0.67cm}   (y=0,1), \\
\label{eqn::strD-sp2} \textsc{SP}_D^1 &\geq& 1-\pr(D'=1 \mid Y=y, Z=1), \quad  (y=0,1).
\end{eqnarray}
The inequalities in \eqref{eqn::strD-sn2} and \eqref{eqn::strD-sp2} are stronger than those in~\eqref{eqn::strD-snsp1}. Therefore, we can omit~\eqref{eqn::strD-snsp1}. 
Solving $0 \leq \pr(Y_0=1 \mid U=c) \leq 1$, we have
\begin{eqnarray*}
\label{proof:thm5:2} &&\pr(Y=1,D'=1 \mid Z=1)-(1-\textsc{SP}_D^1)\pr(Y=1\mid Z=0)-\textsc{SN}_D^1 \textsc{RD}_{Y\mid Z} \geq 0,\\
\label{proof:thm5:3}  &&\pr(Y=0,D'=1 \mid Z=1)-(1-\textsc{SP}_D^1)\pr(Y=0\mid Z=0)+\textsc{SN}_D^1 \textsc{RD}_{Y\mid Z} \geq 0,
\end{eqnarray*}
which imply
\begin{eqnarray}
\label{proof:thm5:4}  && \textsc{SN}_D^1 \leq \frac{\pr(Y=1,D'=1 \mid Z=1)-(1-\textsc{SP}_D^1)\pr(Y=1\mid Z=0)}{\textsc{RD}_{Y\mid Z}},\\
\label{proof:thm5:5}  && \textsc{SN}_D^1 \geq -\frac{\pr(Y=0,D'=1 \mid Z=1)-(1-\textsc{SP}_D^1)\pr(Y=0\mid Z=0)}{\textsc{RD}_{Y\mid Z}}.
\end{eqnarray}
Next, we show that~\eqref{eqn::strD-sn2} and ~\eqref{eqn::strD-sp2} imply~\eqref{proof:thm5:5}. From \eqref{eqn::strD-sp2},
\begin{eqnarray*}
&&\pr(D'=1 \mid Y=0, Z=1)\times \textsc{RD}_{Y\mid Z}\\
&=&\pr(D'=1 \mid Y=0, Z=1) \{\pr(Y=0\mid Z=0)-\pr(Y=0\mid Z=1)\}\\
&=&\pr(D'=1 \mid Y=0, Z=1)\pr(Y=0\mid Z=0)-\pr(Y=0, D'=1  \mid Z=1)\\
&\geq& (1-\textsc{SP}_D^1)\pr(Y=0\mid Z=0)-\pr(Y=0, D'=1  \mid Z=1).
\end{eqnarray*}
Therefore,
\begin{eqnarray*}
&&\pr(D'=1 \mid Y=0, Z=1) \geq -\frac{\pr(Y=0,D'=1 \mid Z=1)-(1-\textsc{SP}_D^1)\pr(Y=0\mid Z=0)}{\textsc{RD}_{Y\mid Z}},
\end{eqnarray*}
which means that ~\eqref{eqn::strD-sn2} implies~\eqref{proof:thm5:5}. As a result, we can omit ~\eqref{proof:thm5:5}.

Combining  ~\eqref{eqn::strD-sn2} with \eqref{proof:thm5:4},
\begin{eqnarray} 
&& \max_{y=0,1}\pr(D'=1 \mid Y=y, Z=1)  \nonumber  \\
 &  \leq & \textsc{SN}_D^1\leq \frac{\pr(Y=1,D'=1 \mid Z=1)-(1-\textsc{SP}_D^1)\times \pr(Y=1\mid Z=0)}{\textsc{RD}_{Y\mid Z}}. \label{eq::uppersnd1}
\end{eqnarray} 
In \eqref{eq::uppersnd1}, the upper bound must be larger than or equal to 0, the lower bound must be smaller than or equal to 1, and the upper bound must be larger than or equal to the lower bound. 
These require
\begin{eqnarray}
&&\frac{\pr(Y=1,D'=1 \mid Z=1)-(1-\textsc{SP}_D^1) \times \pr(Y=1\mid Z=0)}{\textsc{RD}_{Y\mid Z}}
\nonumber  \\
&\geq& \max_{y=0,1}\pr(D'=1 \mid Y=y, Z=1),  \label{eqn::proof:thm5:lowerspd1}  \\
&& \frac{\pr(Y=1,D'=1 \mid Z=1)-(1-\textsc{SP}_D^1)\times \pr(Y=1\mid Z=0)}{\textsc{RD}_{Y\mid Z}} 
 \geq  0.  \label{eqn::proof:thm5:lowerspd2}
\end{eqnarray}
Under $\textsc{RD}_{Y\mid Z}\geq 0$, \eqref{eqn::proof:thm5:lowerspd2} holds. Thus, we can omit \eqref{eqn::proof:thm5:lowerspd2}.  Combining \eqref{eqn::proof:thm5:lowerspd1} with
 ~\eqref{eqn::strD-sp2}, we have
$\textsc{SP}_D^1 \geq 1- \min S_{D}$.

Finally, we derive the bounds on 
\begin{eqnarray}
\label{eqn:str:D}
\CACE  = \frac{(\textsc{SN}_D^1+\textsc{SP}_D^1-1)\textsc{RD}_{Y\mid Z}}{\pr(D'=1 \mid Z=1)-(1-\textsc{SP}_D^1)},
\end{eqnarray} 
where only $D$ is mismeasured.
If $\CACE'\geq 0$, then
from \eqref{eqn:str:D}, 
$\CACE$ is increasing in $\textsc{SN}_D^1$.
Replacing $\textsc{SN}_D^1$ with its bound limits in \eqref{eqn:str:D}, we obtain 
\begin{eqnarray*}
&&\frac{\{\max_{y=0,1}\pr(D'=1 \mid Y=y, Z=1)+\textsc{SP}_D^1-1\}\times \textsc{RD}_{Y\mid Z}}{\pr(D'=1 \mid Z=1)-(1-\textsc{SP}_D^1)} \\
&\leq &  \CACE  \leq  \frac{\pr(Y=1,D'=1\mid Z=1)-(1-\textsc{SP}_D^1)\times \pr(Y=1\mid Z=1) }{\pr(D'=1 \mid Z=1)-(1-\textsc{SP}_D^1)} . \\
\end{eqnarray*}
Because the above bound limits are monotone in $\textsc{SP}_D^1$,  we can obtain the sharp upper and lower bounds on $\CACE$ by replacing $\textsc{SP}_D^1$ with its bound limits.
If  $\pr(D'=1\mid Y =1,Z =1)\geq \pr(D'=1\mid Y =0,Z =1)$, then the bounds  simplify to those in Theorem~\ref{thm:bound:str:D}.

\QEDB

\subsection{Bounds on $\CACE$ under simultaneous measurement errors}
Under strong monotonicity,
it is complicated to obtain closed-form bounds under simultaneous measurement errors of more than one elements of $(Z,D,Y)$. We  propose the general strategy for  calculating the sharp bounds numerically.

We can express $\{ \pr(Z' =z', D'=d',Y'=y')  :   z',d',y'=0,1 \}$ in terms of $\{\pr(Z=z), \pr(U=u),\pr(Y_z=1\mid U=u) : z=0,1; u=n,c\}$ and $(\textsc{SN}'_Z,\textsc{SP}'_Z,\textsc{SN}^1_D,\textsc{SP}^1_D,\textsc{SN}_Y,\textsc{SP}_Y)$.  Therefore, we obtain $8-1$  equality constraints  for $\{\pr(Z=z), \pr(U=u),\pr(Y_z=1\mid U=u) : z=0,1; u=n,c\}$ and $(\textsc{SN}'_Z,\textsc{SP}'_Z,\textsc{SN}^1_D,\textsc{SP}^1_D,\textsc{SN}_Y,\textsc{SP}_Y)$. 
Using linear or non-linear programming, we can numerically calculate the bounds by minimizing and maximizing $\CACE$ under the equality constraints and the inequality constraints 
$
\{ 0\leq  \pr(Z=z) \leq 1, 0\leq \pr(U=u) \leq 1, 0\leq \pr(Y_z=1\mid U=u) \leq 1 : z=0,1; u= n,c \}.
$

\section{Results with differential measurement errors}\label{sec::differentialmeasurementerror}

\noindent {\it Proof of Theorem~\ref{thm:diff:Y}.}  From \eqref{eqn:1} and  \eqref{eqn:2}, 
\begin{eqnarray*}
&&\pr(Y=1\mid Z=1)= \frac{\pr(Y'=1 \mid Z=1)-(1-\textsc{SP}_Y^1)}{\textsc{SN}_Y^1+\textsc{SP}_Y^1-1}, \\
&&\pr(Y=1\mid Z=0)=\frac{\pr(Y'=1 \mid Z=0)-(1-\textsc{SP}_Y^0)}{\textsc{SN}_Y^0+\textsc{SP}_Y^0-1},
\end{eqnarray*}
which implies the formula of   $\CACE $ in Theorem~\ref{thm:diff:Y}.
\QEDB

\vspace{5mm}
\noindent {\it Proof of Theorem~\ref{thm:diff:D}.}  From~\eqref{eqn:1} and~\eqref{eqn:2},  
\begin{eqnarray*}
&&\pr(D=1\mid Z=1)= \frac{\pr(D'=1 \mid Z=1)-(1-\textsc{SP}_D^1)}{\textsc{SN}_D^1+\textsc{SP}_D^1-1}, \\
&& \pr(D=1\mid Z=0)=\frac{\pr(D'=1 \mid Z=0)-(1-\textsc{SP}_D^0)}{\textsc{SN}_D^0+\textsc{SP}_D^0-1},
\end{eqnarray*}
which implies the formula of   $\CACE $ in Theorem~\ref{thm:diff:D}.
\QEDB

%
%

\section{More details for \S \ref{sec::illustration}}\label{sec::illustrationdetails}

 \subsection{Data}
 Table~\ref{tab:data} shows the data in \S 7 in the main text.
 \renewcommand {\thetable} {S\arabic{table}}
 \begin{table}[t]
\caption{Data}\label{tb::data}
\label{tab:data}
\begin{center}
\begin{subtable}{0.5\textwidth}
\caption{Example \ref{eg::1}: \citet{improve2014endovascular}'s study}
\begin{tabular}{cccccc} \hline
      &   \multicolumn{2}{c}{$Z=1$}  &  &  \multicolumn{2}{c}{$Z=0$}  \\  \cline{2-3} \cline{5-6} 
      & $D=1$  &  $D=0$           &       &  $D=1$  &  $D=0$\\
 $Y=1$ &   107    &      68         &      &    24       &    131        \\
 $Y=0$ &   42    &     42            &      &     8         &     79           \\ \hline
\end{tabular}

\label{tab::data1}
\end{subtable}%

\begin{subtable}{0.5\textwidth}
\caption{Example \ref{eg::2}: \citet{hirano2000assessing}'s study}
\begin{tabular}{cccccc} \hline
      &   \multicolumn{2}{c}{$Z=1$}  &  &  \multicolumn{2}{c}{$Z=0$}  \\  \cline{2-3} \cline{5-6} 
      & $D=1$  &  $D=0$           &       &  $D=1$  &  $D=0$\\
 $Y=1$ &   31    &     85         &      &     30       &    99       \\
 $Y=0$ &   424   &     944            &      &     237         &    1041           \\ \hline
\end{tabular}

\label{tab::data2}
\end{subtable}

\begin{subtable}{0.5\textwidth}
\caption{Example \ref{eg::3}: \citet{sommer1991estimating}'s study}
\begin{tabular}{cccccc} \hline
      &   \multicolumn{2}{c}{$Z=1$}  &  &  \multicolumn{2}{c}{$Z=0$}  \\  \cline{2-3} \cline{5-6} 
      & $D=1$  &  $D=0$           &       &  $D=1$  &  $D=0$\\
 $Y=1$ &   9663    &     2385         &      &     0       &    11514       \\
 $Y=0$ &   12   &     34            &      &     0         &    74           \\ \hline
\end{tabular}

\label{tab::data3}
\end{subtable}

\end{center}
\label{default}
\end{table}%

\subsection{A method for constructing confidence intervals for $\CACE$} 
 
 The bounds on $\CACE $ take the form 
\begin{eqnarray*}
&\CACE ' \times l_1 \leq  \CACE  \leq \CACE ' \times u_1,  &\qquad \text{ if } \CACE' \geq 0,\\
&\CACE ' \times l_2 \leq  \CACE  \leq \CACE ' \times u_2, & \qquad \text{ if } \CACE' <0 ,
\end{eqnarray*} 
where $l_1,l_2,u_1$ and $u_2$ are maximums or minimums of the functions of the observed distribution. 
This form of bounds is different from most partially identified parameter in the literature \citep{imbens2004confidence, chernozhukov2013intersection, jiang2018using}. Motivated by \citet{berger1994p}'s method for hypothesis testing, we propose the following strategy for constructing confidence interval.

In the first step, construct $\textsc{CI}'$, a $1-\gamma$ confidence interval for $\CACE '$. In the second step, construct $\textsc{CI}(\tCACE ' )$, a $1-(\alpha-\gamma)$ confidence interval for $\CACE$ when the parameter $\CACE '$ is fixed at the value $  \tCACE '$, for all $ \tCACE ' \in \textsc{CI}'$. In the third step, construct the final confidence interval by taking the union of these $\textsc{CI}(  \tCACE ')$'s:
$$
\text{CI} = \cup_{   \tCACE   ' \in \text{CI}'}\textsc{CI}(   \tCACE   ') .
$$ 

This strategy is easy to implement. In the first step, we can construct $\textsc{CI}'$ based on standard techniques. In the second step, we can construct $\textsc{CI}(\tCACE ' )$ using the method of \citet{imbens2004confidence} or \citet{jiang2018using} for partially identified parameters with interval bounds.

We then prove that this confidence interval has a coverage rate at least as large as $1-\alpha$.   

 {\it Proof. } The conclusion follows from 
 \begin{eqnarray*}
&&\pr( \CACE \notin \text{CI} )\\
&=& \pr\left\{  \CACE \notin \cup_{ \tCACE ' \in \text{CI}'}\textsc{CI}( \tCACE '),  \CACE ' \in \textsc{CI}' \right\}
+\pr\left\{  \CACE \notin \cup_{\tCACE ' \in \text{CI}'}\textsc{CI}(\tCACE '),  \CACE ' \notin \textsc{CI}' \right\} \\
&\leq& \pr\left\{  \CACE \notin \cup_{\tCACE ' \in \text{CI}'}\textsc{CI}(\tCACE '),  \CACE ' \in \textsc{CI}' \right\}
+\pr(\CACE ' \notin \textsc{CI}')\\
&\leq &\pr\left\{  \CACE \notin \textsc{CI}(\CACE ') \right\} +\gamma\\
&=&\alpha-\gamma+\gamma\\
&=&\alpha. 
\end{eqnarray*}
The proof above is based on finite-sample exact confidence intervals. It carries over to large-sample confidence intervals.
\QEDB

 \section{Other results}\label{sec::otherresults}
  \subsection{More discussion on the non-differential measurement error of $Z$}
  Let $\widehat{\textsc{RD}}_{Y\mid Z}$ and $\widehat{\textsc{RD}}_{D\mid Z}$ be the estimators of $\textsc{RD}_{Y\mid Z}$  and $\textsc{RD}_{D\mid Z}$, respectively. 
Without measurement error, applying the central limit theorem, we have 
\begin{eqnarray*}
n^{1/2} \begin{pmatrix}
\widehat{\textsc{RD}}_{Y\mid Z}\\
\widehat{\textsc{RD}}_{D\mid Z}
\end{pmatrix}  \longrightarrow  \mathcal{N}_2  \left\{  
\begin{pmatrix}
\textsc{RD}_{Y\mid Z}\\
\textsc{RD}_{D\mid Z}
\end{pmatrix},   \begin{pmatrix}
\sigma_1^2 & \rho \sigma_1\sigma_2\\
\rho \sigma_1\sigma_2 & \sigma_2^2
\end{pmatrix}
\right\}
\end{eqnarray*}
in distribution, where $\sigma_1^2, \sigma_2^2, \rho \sigma_1\sigma_2$ are the asymptotic variances and covariance of  $n^{1/2} \widehat{\textsc{RD}}_{Y\mid Z}$  and $n^{1/2} \widehat{\textsc{RD}}_{Y\mid Z}$, respectively.
Using the delta method, we obtain that the asymptotic variance of the naive estimator $\widehat{\textsc{RD}}_{Y\mid Z} /  \widehat{\textsc{RD}}_{D\mid Z}$ is 
$
(\sigma_1^2 - \CACE  \rho\sigma_1\sigma_2 +\CACE ^2\sigma^2_2   ) / \textsc{RD}^2_{D\mid Z}.
$

When $Z$ is mismeasured, let $r'_Z=\textsc{SN}'_Z+\textsc{SP}'_Z-1$. From Theorem~\ref{thm:cace},  $\textsc{RD}_{Y\mid Z'}= r'_Z \cdot \textsc{RD}_{Y\mid Z}$ and  $\textsc{RD}_{D\mid Z'}= r'_Z \cdot \textsc{RD}_{D\mid Z}$. Therefore,
\begin{eqnarray*}
n^{1/2}  \begin{pmatrix}
\widehat{\textsc{RD}}_{Y\mid Z ' }\\
\widehat{\textsc{RD}}_{D\mid Z ' }
\end{pmatrix}  \longrightarrow  \mathcal{N}_2  \left\{ 
\begin{pmatrix}
r'_Z\cdot\textsc{RD}_{Y\mid Z}\\
r'_Z\cdot\textsc{RD}_{D\mid Z}
\end{pmatrix},   (r'_Z)^2  \begin{pmatrix}
\sigma_1^2 & \rho \sigma_1\sigma_2\\
\rho \sigma_1\sigma_2 & \sigma_2^2
\end{pmatrix}
\right\} ,
\end{eqnarray*}
in distribution. 
Using the delta method, we obtain the  asymptotic variance of the naive estimator $\widehat{\textsc{RD}}_{Y\mid Z ' }  / \widehat{\textsc{RD}}_{D\mid Z ' }$ is
$
( \sigma_1^2 - \CACE  \rho\sigma_1\sigma_2 +\CACE ^2\sigma^2_2  ) / \textsc{RD}^2_{D\mid Z}.
$
 Therefore, the non-differential measurement error of $Z$ does not affect the asymptotic variance of the naive estimator.

 \subsection{Dichotomization of a discrete treatment}
 \setcounter{assumption}{0}
\renewcommand {\theassumption} {S\arabic{assumption}}

We show that $ \tau_{\text{2sls}} = \tau_{\text{2sls}} '  \times w_k  $, where 
$w_k = \pr(D_1  \geq k>D_0) / \sum_{j=1}^J \pr(D_1  \geq j>D_0)  $ if Assumptions \ref{asm:iv}(a) and (b) hold. This follows because
\begin{eqnarray*}
\tau_{\text{2sls}}&=&  \tau_{\text{2sls}} ' \times 
\frac{    \E(D' \mid Z=1)- \E(D' \mid Z=0)   }{  \E(D \mid Z=1)- \E(D \mid Z=0)  } \\
&=&  \tau_{\text{2sls}} ' \times  \frac{\pr(D_1 \geq k )-\pr(D_0 \geq k )}{  \sum_{j=1}^J   \{  \pr(D_1 \geq j )-\pr(D_0 \geq j ) \} } \\
&=& \tau_{\text{2sls}} ' \times  \frac{ \pr(D_1  \geq k>D_0) }{ \sum_{j=1}^J     \pr(D_1 \geq j  > D_0   )   } \\
&=& \tau_{\text{2sls}} '  \times w_k.
\end{eqnarray*}

\pdfbookmark[1]{References}{References}
\spacingset{1.45}
\bibliographystyle{Chicago}
\bibliography{paper-ref}

\end{document}